%% file: main.tex
\documentclass[fleqn,usenatbib]{mnras}

\input{header}

\title[The anisotropic density variance]{A multi-shock model for the density variance of anisotropic, highly-magnetised, supersonic turbulence}

\author[Beattie, Mocz, Federrath \& Klessen]{James R. Beattie$^{\orcidicon{0000-0001-9199-7771}\,1}$\thanks{E-mail: james.beattie@anu.edu.au}, 
Philip Mocz$^{\orcidicon{0000-0001-6631-2566}\,3,\dagger}$,
Christoph Federrath$^{\orcidicon{0000-0002-0706-2306}\,1,2}$ and 
\newauthor{Ralf S. Klessen$^{\orcidicon{0000-0002-0560-3172}\,4,5}$}
\\
$^{1}$Research School of Astronomy and Astrophysics, Australian National University, Canberra, ACT 2611, Australia \\
$^{2}$Australian Research Council Centre of Excellence in All Sky Astrophysics (ASTRO3D), Canberra, ACT 2611, Australia \\
$^{3}$Department of Astrophysical Sciences, Princeton University, 4 Ivy Lane, Princeton, NJ 08544, USA \\
$^{4}$Universit\"at Heidelberg, Zentrum f\"ur Astronomie, Institut f\"ur Theoretische Astrophysik, Albert-Ueberle-Str. 2, 69120 Heidelberg, Germany\\ 
$^{5}$Universit\"at Heidelberg, Interdisziplin\"ares Zentrum f\"ur Wissenschaftliches Rechnen, Im Neuenheimer Feld 205, 69120 Heidelberg, Germany \\
$^{\dagger}$Einstein Fellow 
}

\date{Accepted XXX. Received YYY; in original form ZZZ}

\pubyear{2020}

\begin{document}
\label{firstpage}
\pagerange{\pageref{firstpage}--\pageref{lastpage}}
\maketitle

\begin{abstract}
Shocks form the basis of our understanding for the density and velocity statistics of supersonic turbulent flows, such as those found in the cool interstellar medium (ISM). The variance of the density field, $\sigma^2_{\rho/\rho_0}$, is of particular interest for molecular clouds (MCs), the birthplaces of stars in the Universe. The density variance may be used to infer underlying physical processes in an MC, and parameterises the star formation (SF) rate of a cloud. However, models for $\sigma^2_{\rho/\rho_0}$ all share a common feature -- the variance is assumed to be isotropic. This assumption does not hold when a trans/sub-Alfv\'enic mean magnetic field, $\vecB{B}_0$, is present in the cloud, which observations suggest is relevant for some MCs. We develop an anisotropic model for $\sigma_{\rho/\rho_0}^2$, using contributions from hydrodynamical and fast magnetosonic shocks that propagate orthogonal to each other. Our model predicts an upper bound for $\sigma_{\rho/\rho_0}^2$ in the high Mach number $(\M)$ limit as small-scale density fluctuations become suppressed by the strong $\vecB{B}_0$. The model reduces to the isotropic $\sigma_{\rho/\rho_0}^2-\M$ relation in the hydrodynamical limit. To validate our model, we calculate $\sigma_{\rho/\rho_0}^2$ from 12~high-resolution, three-dimensional, supersonic, sub-Alfv\'enic magnetohydrodynamical (MHD) turbulence simulations and find good agreement with our theory. We discuss how the two MHD shocks may be the bimodally oriented over-densities observed in some MCs and the implications for SF theory in the presence of a sub-Alfv\'enic $\vecB{B}_0$. By creating an anisotropic, supersonic density fluctuation model, this study paves the way for SF theory in the highly anisotropic regime of interstellar turbulence. 
\end{abstract}


\section{Introduction}\label{sec:intro}
    Shocks are the fundamental building blocks of supersonic turbulence and are key to understanding both the density and velocity statistics of these flows which are ubiquitous in many astrophysical phenomena \citep{Burgers1948,Vazquez1994,Padoan1997,Klessen2000,Padoan2011,Federrath2013,Lehmann2016,Robertson2018,Mocz2018,Park2019,Abe2020,Federrath2021}. Density statistics are of particular interest for understanding the structure and physical processes that shape and govern the dynamics of the compressible, supersonic interstellar medium (ISM;  \citealt{MacLow2004,Krumholz2005,Federrath2008,Federrath2009,Burkhart2009,Padoan2011,Hennebelle2011,Federrath2012,Burkhart2012,Schneider2012,Konstandin2012,Schneider2013,Burkhart2014,Klessen2016,Burkhart2018,Mocz2018,Mocz2019,Beattie2020,Menon2020}). One such important statistic is the density probability density function, the $\rho\,$-PDF. For example, simple models of molecular clouds (MCs) in the ISM, which are supersonic and magnetised, and have not yet started to collapse under their own self-gravity have an \textit{approximately} log-normal volume-weighted $\rho\,$-PDF,
    \begin{align}
        p_s(s;\sigma_s^2)\d{s} &= \frac{1}{\sqrt{2\pi\sigma_s^2}} \exp \left( -\frac{(s- s_0)^2}{2\sigma_s^2} \right)\d{s}, \label{eq:dis} \\
        s &\equiv \ln(\rho / \rho_0), \\
        s_0 &= - \frac{\sigma_s^2}{2}, \label{eq:vars_0 relation}\\
        \sigma^2_s &= f(\M,\Ma,b,\gamma,\Gamma,\ell_{\rm D}), \label{eq:var}
    \end{align}
    where $\rho$ is the cloud density, with $\rho_0$ being its volume-weighted mean value. The log-density variance, $\sigma_s^2$, is the key parameter for this model. It is a function of ({i}) the turbulent Mach number, 
    \begin{align} \label{eq:M}
    \M = \sigma_{V}/c_s,
    \end{align}
    where $\sigma_{V}$ is the root-mean-square (rms) velocity and $c_s$ is the sound speed \citep{Vazquez1994,Padoan1997,Passot1998,Price2011,Konstandin2012a}, 
    ({ii}) the Alfv\'en Mach number, 
    \begin{align}\label{eq:Ma}
        \Ma = \sigma_{V} / V_{\rm A} = c_s \M / V_{\rm A},
    \end{align}
    where $V_\text{A} = B / \sqrt{4\pi\rho}$, is the Alfv\'en wave velocity and $B$ is the magnetic field \citep{Padoan2011,Molina2012}, 
    ({iii}) the turbulent driving parameter $b$ \citep{Federrath2008,Federrath2010}, 
    ({iv}) the adiabatic index $\gamma$ \citep{Nolan2015}, 
    ({v}) the polytropic index $\Gamma$ \citep{Federrath2015}, and 
    ({vi}) the turbulence driving scale $\ell_{\rm D}$, i.e., the scale on which energy is injected into the system \citep{Bialy2020}. According to this model, the density variance plays two important roles: First, it encapsulates all of the physical processes that influence the density fluctuations of a cloud, fully parameterising the log-normal density fluctuation theory, and
    second, the density variance is the key ingredient for turbulence-regulated star formation theories, including for both determining the star formation rate of an MC, directly through integrating the $\rho$-PDF and for setting the width of the Gaussian component of the initial core and stellar mass function,   \citep[e.g.][]{Padoan2002,Hennebelle2008,Hennebelle2009,Krumholz2005,Padoan2011,Hennebelle2011,Federrath2012,Hopkins2013a,Kainulainen2014,Burkhart2018,Krumholz2019}.
    
    However, all of these studies explicitly (or implicitly) assume that the variance is distributed isotropically in space, i.e., that the magnetic, density or velocity fluctuations do not have a preferential direction. This assumption becomes especially violated in the presence of strong magnetic mean fields creates significant anisotropy in the flow which changes how the momentum and thus the energy is transported along and across the mean magnetic field, $\vecB{B}_0$ \citep[for incompressible, sub-Alfv\'enic turbulence see][]{Goldreich1995,Cho2002,Boldyrev2006,Skalidis2020}. 
    
    Nature seems to be more than content with exploring this anisotropic regime in the ISM. In a detailed analysis of the velocity gradients \citet{Hu2019} finds a number of star-forming MCs that are trans- to sub-Alfv\'enic, i.e. $\Mao \lesssim 1 \implies \rho_0 \sigma_V^2 \lesssim B_0^2$, where $\Mao$ is the Alfv\'en Mach number of the mean-field\footnote{\label{foot:Ma0}We use the same definition as Equation~\ref{eq:Ma}, but using the mean magnetic field instead of the total field, hence, $\Mao = \sigma_V / V_{\rm A0}$, where $V_{\rm A0} = B_0 / \sqrt{4\pi\rho_0}$. }. Hence in the sub-Alfv\'enic mean-field regime the magnetic energy is greater than or within equipartition of the turbulent energies. \citet{Hu2019} reported upon Taurus ($\Mao = 1.19 \pm 0.02$), Perseus~A ($1.22 \pm 0.05$), L1551 ($0.73 \pm 0.13$), Serpens ($0.98 \pm 0.08$), and NGC1333 ($0.82 \pm 0.24$)\footnote{Not all MCs are sub-Alfv\'enic in nature. See for example \citet{Lunttila2008} for quite convincing results that clouds observed in \citet{Troland2008} are super-Alfv\'enic based on the magnetic and column density field correlations. The clouds we list as sub-Alfv\'enic are not part of the survey reported in \citet{Troland2008}.}. Another extremely magnetised cloud is the central molecular zone cloud, G0.253+0.016, studied in \citet{Federrath2016b}. \citet{Pillai2015} measures a mean field of $B_0 = (2.07 \pm 0.95) \,\text{mG}$ for this cloud, and with the density and velocity quantities measured in \citet{Federrath2016b} we find a corresponding Alfv\'en Mach number, $\Mao = 0.3 \pm 0.2$, placing it well within this anisotropic regime. For more details of this calculation, see \citet{Beattie2020c}. 
    
    Furthermore, recent analyses of both CO and column density maps of the Taurus cloud hint that sub-Alfv\'enic flows are present, simply based upon the observed density structures in the clouds. For example, in the $^{13}$CO map of the Taurus cloud \citet{Palmeirim2013} and \citet{Heyer2020} find bimodal distributions of density orientations with respect to the plane-of-sky magnetic field, which have been associated with sub-Alfv\'enic turbulence \citep{Li2013,Soler2013,Burkhart2014,Soler2017,Soler2017a,Tritsis2018,Beattie2020,Seifried2020,Kortgen2020}. We discuss these oriented density structures further, and in much more detail in~\S\ref{sec:discussion}. Hence, to understand star formation and the physical processes that create density fluctuations in this extremely magnetised, supersonic turbulence regime, one must first construct a model for $\sigma_{\rho/\rho_0}^2$, which is the primary goal of this study. 
    
    This study is organised as follows: in \S\ref{sec:densityPDFs}, we consider the underlying motivation for the $\rho$-PDF model in the context of turbulent fluctuations and the central limit theorem from the pioneering work of \citet{Vazquez1994}. In \S\ref{sec:shockModels} we revisit the \citet{Molina2012} model for the isotropic volume-weighted variance using planar shocks and generalise it to include contributions from two different types of shocks: hydrodynamical and fast magnetosonic, which propagate orthogonal to each other. In \S\ref{sec:numerics} we outline the anisotropic, supersonic MHD turbulence simulations that we use to test our new anisotropic density variance model on. In \S\ref{sec:varianceModel} we discuss shock volume-filling fractions and the fitting procedure for our two-shock variance model. In \S\ref{sec:results} we introduce results for the fit to the 3D density variance data, and explore the limiting behaviour of the model. In \S\ref{sec:discussion} we discuss implications that our study has for oriented density structures and star formation theory in highly-magnetised environments of the ISM, and finally in \S\ref{sec:conclusion} we summarise and itemise our key results.

    \section{The density PDF and variance}\label{sec:densityPDFs}
    
    \subsection{Log-normal models}\label{sec:logNormal}
    
    Log-normal models for the density can be traced back to \citet{Vazquez1994}, where they explore different functional forms for the $\rho\,$-PDF in order to model the self-similar, hierarchical structure of density in the cool, pressureless $(\M \gg 1)$, non-self-gravitating ISM regime. \citet{Vazquez1994} considers the density fluctuation random variable, $\delta = \rho / \rho_0$, and motivates the log-normal PDF by assuming that for some time, $t_n$, the density can be expressed as a multiplicative interaction through independent density fluctuations, hence,
    \begin{align}
        \rho(t_n) = \delta_n \delta_{n-1} \hdots \delta_1 \delta_0 \rho(t_0) = \left(\prod^n_{i=0} \delta_i\right) \rho(t_0),
    \end{align}
    where $\rho(t_0)$ is the initial density. Under the log-transform, the product becomes a sum,
    \begin{align}
        \ln\rho(t_n) = \left(\sum^n_{i=0} \ln\delta_i\right) + \ln\rho(t_0).
    \end{align}    
    If each $\delta$ follows the same underlying distribution and are independent from each other (i.e., not correlated) then the central limit theorem states that the
    distribution of the log-density affected by this additive random process is approximately normal, specifically:
    \begin{align}
     \sqrt{n}\left[ \ln \rho(t_n) - \Exp{\ln\rho}_t \right] = \sqrt{n} s_n \rightarrow \mathcal{N}(0,\sigma^2_{s}),
    \end{align}
    where $\mathcal{N}(0,\sigma_s^2)$ is the normal distribution with mean zero, as $n$, the number of density fluctuations, approaches infinity. This lets us understand the nature of a single density fluctuation changing in time. However, \citet{Vazquez1994} further argues that since the hydrodynamical equations are self-similar in space (i.e., invariant under arbitrary length scalings), the fluctuations should be log-normal globally. \citet{Passot1998} extend this idea, recasting $\delta$ as a perturbation in the density from a transient shock, rather than just a generic turbulent fluctuation. This means that one could relate the well-known shock-jump relations to the $\sigma_{\rho}$ variable\footnote{For a log-normally distributed variable, $\rho/\rho_0$, the variance of the log-variable obeys $\sigma_{\ln\rho/\rho_0}^2 = \ln(1 + \sigma^2_{\rho/\rho_0})$ and is the value that is commonly reported in the literature. Throughout this study we however consider $\sigma^2_{\rho/\rho_0}$, without making any assumptions of the underlying distribution.}, formulating the relation,
    \begin{align}
        \sigma_{\rho/\rho_0} \propto \M.
    \end{align}
    As described in \S\ref{sec:intro}, the density $ \sigma^2_{\rho/\rho_0} - \M$ relation has been studied intensely over the last two decades. The main idea is that $\sigma^2_{\rho/\rho_0}$, or the spread of the $\rho$-PDF is a function of the underlying physical processes that govern the fluid, whether it be motivated by the supersonic plasmas of the ISM  \citep[e.g][]{MacLow2004,Federrath2008,Federrath2010,Price2011,Padoan2011,Federrath2012,Ginsburg2013,Federrath2015,Klessen2016} or the subsonic density fluctuations in the hot, stratified, intracluster medium (ICM; e.g. \mbox{\citealt{Mohapatra2020b,Mohapatra2020}}). More specifically, for an isothermal, turbulent, hydrodynamic medium, 
    \begin{align}
        \sigma_s^2 = \ln(1 + b^2\M^2),
    \end{align}
    where $b$ is controlled by the amount of solenoidal ($\nabla \times \vecB{F}$) and compressive ($\nabla \cdot \vecB{F}$) modes injected by the turbulence driving field $\vecB{F}$ \citep{Federrath2008,Federrath2010,Price2011}. Introducing a non-isothermal equation of state gives the relation,
    \begin{align}
        \sigma_s^2 = 
        \left\{
        \begin{matrix}
            \ln(1 + b^2\M^{(5\gamma + 1)/3}), &b\M \leq 1, \\ 
            \ln\left( 1 + \dfrac{(\gamma + 1)b^2\M^2}{(\gamma - 1)b^2\M^2 + 2}  \right), &b\M > 1,
        \end{matrix}
        \right.
    \end{align}
    where $\gamma$ is the adiabatic index of the medium \citep{Nolan2015}. For isothermal and magnetised turbulence, the variance changes with the correlation between $B$ and $\rho$,
    \begin{align} \label{eq:sigma_magnetic}
        \sigma_s^2 &= 
        \left\{
        \begin{matrix}
            \ln\left( 1 + b^2\M^2\dfrac{\beta_0}{ \beta_0 + 1}\right),   & B \propto \rho^{1/2}, \\ 
            \ln\left( 1 + \dfrac{1}{2}\left[ \sqrt{\left(1 + \beta_0 \right)^2 + 4b^2\beta_0 } \right] -1 - \beta_0 \right),  & B \propto \rho, 
        \end{matrix}
        \right.
    \end{align}
    where $\beta_0 = 2 \MaO{2}/\M^2$ is the plasma beta with respect to the mean magnetic field, which was explored in \citet{Molina2012} and will be discussed in more detail in \S\ref{sec:shockGeometry}.

    \subsection{The relation between shocks and the density variance}

     The relation between density contrasts and the volume-weighted variance of the underlying field is given by
    \begin{align} \label{eq:densVariance}
        \sigma_{\rho/\rho_0}^2 &= \frac{1}{\V} \int_{\V} \d{\V} \, \left( \frac{\rho}{\rho_0} - \Exp{\frac{\rho}{\rho_0}} \right)^2 = \frac{1}{\V} \int_{\V} \d{\V} \, \left( \frac{\rho}{\rho_0} - 1 \right)^2,
    \end{align}
    where $\V$ is the volume for the fluid region of interest \citep{Padoan2011}. \citet{Molina2012} turn Equation~\ref{eq:densVariance} into an integral over density contrasts by constructing the function $\d{\V} = f(\rho_1 / \rho_0) \d{(\rho_1 / \rho_0)}$ based on the shock geometry, where $\rho_1 / \rho_0$ is the density contrast for a single shock. \citet{Molina2012} use this to derive an analytical model for $\sigma^2_{\rho/\rho_0}$ for different $B - \rho$ correlations, assuming that the shock geometry is the same for all shocks in the magnetised plasma, and that there is no preferential direction for the density contrast produced by the shocks, i.e., that they are isotropic. This is a good approximation for $\Mao \geq 2$, when the turbulent component of the magnetic field is larger than or equal to the mean-field component \citep{Beattie2020c}. However, it breaks down for $\Mao \lesssim 1$, i.e., when the mean field is very strong, relevant to this study.
 
    Anisotropic, sub-Alfv\'enic, compressive density structures were studied in detail in \citet{Beattie2020}, showing that the anisotropy of the density fluctuations is a function of both $\M$ and $\Mao$. They attributed the anisotropy to hydrodynamical shocks that form perpendicular to $\vecB{B}_0$, causing parallel fluctuations, and fast magnetosonic waves that form parallel to $\vecB{B}_0$, causing perpendicular fluctuations, illustrated schematically in Figure~\ref{fig:shock_schematic}. This is consistent with findings from \citet{Tritsis2016}, who showed that observations of striations (density perturbations that form parallel to $\vecB{B}_0$ in sub-Alfv\'enic flows) are reproducible when one considers fast magnetosonic wave perturbations. In this study, we take this phenomenology further and model the variance of density structures arising from a supersonic velocity field with a strong magnetic guide field, considering these two types of shocks. 
    
    The general form of this variance can be constructed as follows\footnote{By expressing the variance as we have below we are assuming that the density fluctuations are symmetric around the magnetic field. This was shown explicitly through the elliptic symmetry with respect to the magnetic field of the 2D density power spectra in Figure~2 of \citet{Beattie2020} and 2D structure functions in \citet{Hu2020b}.},
    \begin{align}
        \sigma^2_{\rho/\rho_0} = \sigma^2_{\rho/\rho_0 \parallel} + \sigma^2_{\rho/\rho_0 \perp} + 2\sigma_{\rho/\rho_0 \parallel}\sigma_{\rho/\rho_0 \perp},
    \end{align}
    which decomposes the total variance into components of the density variance parallel to $\vecB{B}_0$, termed $\sigma^2_{\rho/\rho_0 \parallel}$, and perpendicular to $\vecB{B}_0$, termed $\sigma^2_{\rho/\rho_0 \perp}$. By assuming that these fluctuations are independent from one another we can further simplify the equation to
    \begin{align}
        \sigma^2_{\rho/\rho_0} \approx \sigma^2_{\rho/\rho_0 \parallel} + \sigma^2_{\rho/\rho_0 \perp},
    \end{align}
    where the correlation term, $2\sigma_{\rho/\rho_0 \parallel}\sigma_{\rho/\rho_0 \perp}$, between the fluctuations becomes zero in the statistical average\footnote{This is a justified assumption because the statistically-averaged elliptic power spectra in Figure~2 of \citet{Beattie2020} do not show any rotations in the $k_{\perp} - k_{\parallel}$ plane, i.e. the principal axes of the ellipses fitted to the power spectra are aligned with the $k_{\perp} - k_{\parallel}$ coordinate axis, where $k_{\perp}$ corresponds to wave numbers perpendicular to $\vecB{B}_0$ and $k_{\parallel}$ for parallel wave numbers. This means that the density fluctuations along and across the mean field are statistically independent from each other, and hence $2\sigma_{\rho/\rho_0 \parallel}\sigma_{\rho/\rho_0 \perp} \approx 0$.}. For a short discussion on how the velocity and magnetic field is arranged in this turbulence regime we refer the reader to Appendix~\ref{app:orientation}. The main purpose of this study is therefore to explore and model each of these components, extending the work of \citet{Molina2012} and \citet{Beattie2020} and indeed the numerous works on $\sigma^2_{\rho/\rho_0}$, into the sub-Alfv\'enic, anisotropic, supersonic turbulent regime. Furthermore, with the development of a density fluctuation model, this is the first step towards an anisotropic, magneto-turbulent star formation theory.
 
\section{Deriving an anisotropic density variance model}\label{sec:shockModels}

    In this section we construct an anisotropic variance model. We first explore some geometrical features of shocks, summarising the key works of \citet{Padoan2011} and \citet{Molina2012}, and derive Equation~\ref{eq:densVariance}. Next we create our two-shock model for the anisotropic density field, based on hydrodynamical shocks that travel parallel to the mean magnetic field, which we call type~I shocks, and fast magnetosonic shocks that travel perpendicular to the mean magnetic field, which we call type~II shocks. Finally, we discuss the volume-filling fractions of the two shock types and the limiting behaviour of the model. 

    \subsection{Geometry of a shock}\label{sec:shockGeometry}

    Consider a planar shock with surface area $\ell^2_{\rm shock}$ and shock width $\lambda$, propagating in a system with volume $\mathcal{V} = L^3$. The volume of the shock is then 
    \begin{align} \label{eq:shockVol}
        \V_{\text{shock}}= \lambda L^2.
    \end{align}
    The shock width is proportional to the density jump between the pre-shock ($\rho_0$) and post-shock ($\rho_1$) densities, multiplied by the integral scale of the turbulence, $\theta L$, i.e., the scale where velocity structure is no longer correlated,
    \begin{align}\label{eq:shockThick}
        \lambda \approx \theta L \frac{\rho_0}{\rho_1},
    \end{align}
    which is derived in \citet{Padoan2011} by balancing the ram and thermal pressures for a shock in hydrodynamical turbulence. We substitute Equation~\ref{eq:shockThick} into \ref{eq:shockVol} to reveal the volume of the shock in terms of the density contrast,
    \begin{equation}
        \V_{\text{shock}}\approx \theta L^3 \frac{\rho_0}{\rho_1}.
    \end{equation}
    Now assuming that $\d{\V} \approx \d{\V_{\text{shock}}}$ we can construct the volume differential, substitute it into Equation~\ref{eq:densVariance} and integrate it to construct the variance as a function of density contrasts. Hence,\footnote{We consider the transformation $\d\V = |\frac{\d\V}{\d(\rho_1/\rho_0)} |\d(\rho_1/\rho_0)$ to simplify the treatment of the limits in Equation~\ref{eq:volume}.}
    \begin{align} \label{eq:dVdRho}
        \d{\V} & \approx \theta L^3 \left( \frac{\rho_0}{\rho_1} \right)^2 \d{\left(\frac{\rho_1}{\rho_0}\right)},
    \end{align}
    and therefore we can rewrite Equation~\ref{eq:densVariance} in terms of density contrasts between $1 \leq \rho_1 / \rho_0 \leq \rho / \rho_0$\footnote{Note that we integrate $\rho_1 / \rho_0$ from 1, i.e., when the pre- and post-shock densities are the same, up to an arbitrarily large density contrast, $\rho / \rho_0$. This is the range of density contrasts where a shock is well defined.}, which is the density domain for which shocks are well-defined in,
    \begin{align} 
        \sigma_{\rho/\rho_0}^2 &\approx \frac{1}{\V} \int_{1}^{\rho / \rho_0} \d{\left(\frac{\rho_1}{\rho_0}\right)} \, \V \theta\left( \frac{\rho_0}{\rho_1} \right)^2  \left( \frac{\rho_1}{\rho_0} - 1 \right)^2, \label{eq:volume}\\
        &= \theta\int_{1}^{\rho / \rho_0} \d{\left(\frac{\rho_1}{\rho_0}\right)} \, \left( 1 - \frac{\rho_0}{\rho_1} \right)^2, \\
        \frac{\sigma_{\rho/\rho_0}^2}{\theta} &\approx \underbrace{\frac{\rho}{\rho_0}}_{\mathclap{\text{over-densities}}} - \overbrace{\frac{\rho_0}{\rho}}^{\mathclap{\text{under-densities}}} - \underbrace{2\ln\left( \frac{\rho}{\rho_0} \right)}_{\mathclap{\substack{\text{logarithmic} \\ \text{over-densities}}}}, \label{eq:varDerv}
    \end{align}
    where $\V = L^3$ cancels between the differential element of the shock and volume-weighted integral. What this means is that for a single-shock model we average over the shock-jump conditions through the whole volume of interest. This will be important for generalising this model to multiple shocks in \S\ref{sec:twoShockModel}. The first term in Equation~\ref{eq:varDerv}, $\rho/\rho_0$, the over-densities, dominates the total variance, since the amplitude of the density fluctuations is large in supersonic turbulence. Both $\rho_0/\rho$, the under-densities, and $\ln(\rho/\rho_0)$, the logarithmic over-densities are small in comparison, allowing us to approximate the total variance as solely dependent upon the over-densities \mbox{\citep{Padoan2011}}. Assuming the integral scale factor is $\theta \approx 1$, i.e., of order the system scale \citep{Federrath2012}, and making the above approximation we find the relation,
    \begin{align}\label{eq:DenVarShockJump}
        \sigma_{\rho/\rho_0}^2 \approx \frac{\rho}{\rho_0},
    \end{align}
    which is the key result that \citet{Padoan2011} and \citet{Molina2012} use to relate the variance to the shock-jump conditions. Before generalising this result, we first consider the details of the two shock types that we use to construct our model of the anisotropic density variance.
    
    \subsection{Shocks in MHD turbulence}
    We have now seen that the variance of a stochastic density field full of shocks can be related to the shock-jump conditions. The shock-jump conditions can be derived in the regular Rankine-Hugoniot fashion, where we equate the upstream and downstream $\vecB{B}$, $\rho\vecB{v}$ and $\rho$ (for an isothermal shock) across the shock boundary, in the local frame of the shock, to conserve energy, momentum and mass \citep{Landau1959}, and we follow \citet{Molina2012} to include the magnetic pressure contribution in the derivation. In the following two subsections we will describe two different types of shocks that are derived using this method.
    
    \begin{figure}
        \centering
        \includegraphics[width=0.9\linewidth]{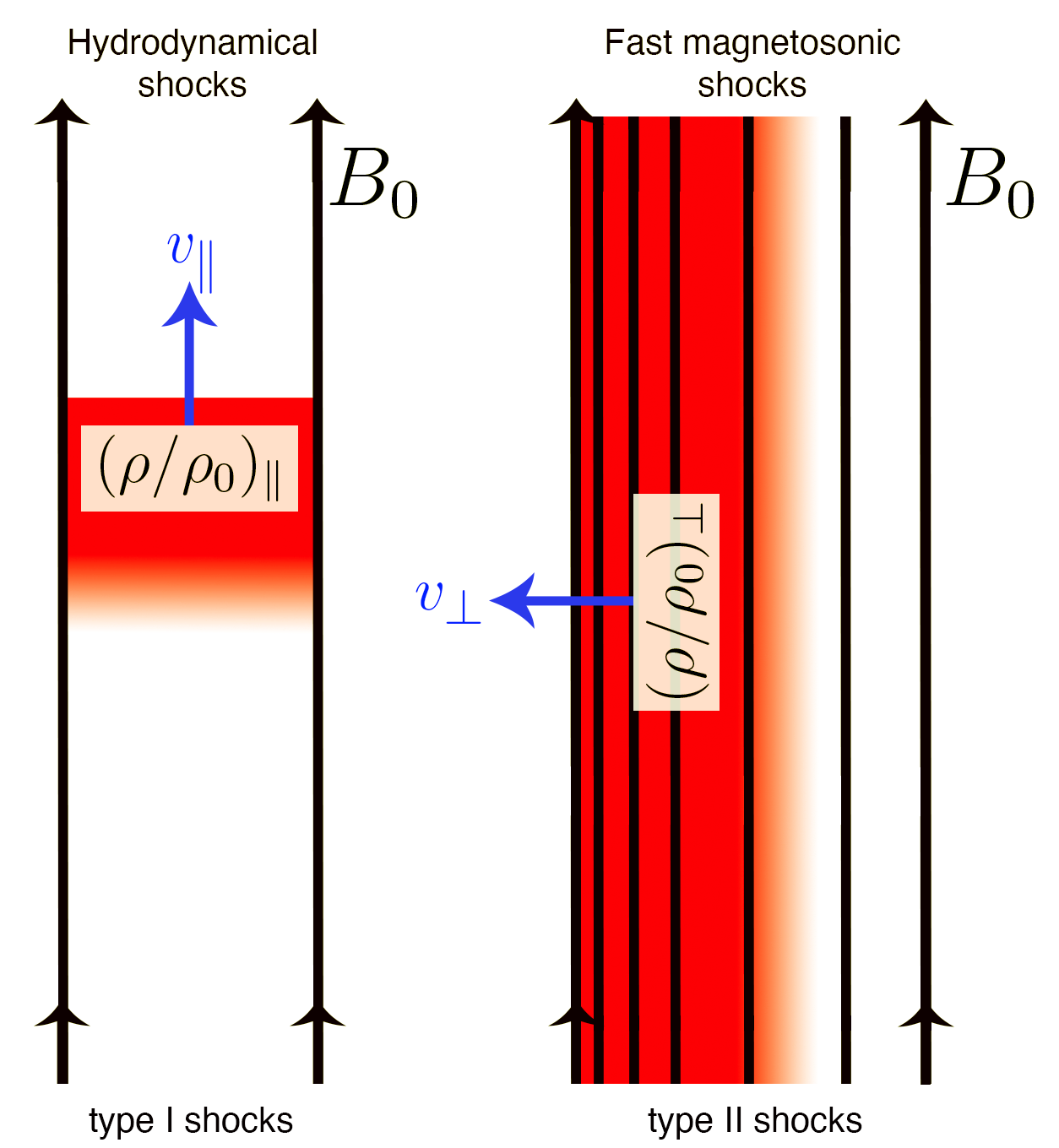}
        \caption{A schematic of the orientation, propagation direction and shock jump, $\rho/\rho_0$, of the two shocks discussed in \S\ref{sec:shockModels}. \textbf{Left:} type~I, hydrodynamical shocks form from streaming material up and down $\vecB{B}_0$. \textbf{Right:} type~II, fast magnetosonic shocks that form from longitudinal compressions of the magnetic field lines. These cause longitudinal compressions in the density field because the density is flux-frozen to the magnetic field.}
        \label{fig:shock_schematic}
    \end{figure}     
    
    \subsubsection{Type I: hydrodynamic shocks}\label{sec:HDplanarshock}

    In isotropic, supersonic turbulence shocks are randomly oriented in space. However, in the presence of a strong mean magnetic field, where $\delta B \ll B_0$, shocks form from velocity gradients along the magnetic field \citep{Beattie2020c}. We call these shocks type~I shocks. The shocks create large density contrasts perpendicular to the mean field that propagate parallel to it \citep{Beattie2020}. Since the propagation is parallel to $\vecB{B}_0$ the shocks do not feel the Lorentz force (nor does the magnetic field feel the shock, with upstream and downstream $\vecB{B}$ remaining unchanged), and hence the shock-jump conditions are hydrodynamical in nature \citep{Mocz2018}. This is the supersonic (non-linear) equivalent of slow or acoustic modes from linearised MHD turbulence theory. For an isothermal fluid, with turbulence driving parameter $b$, the shock jump is
    \begin{align}\label{eq:hdJump}
        \left(\frac{\rho}{\rho_0}\right)_{\parallel} = b^2 \M^2,
    \end{align}
    where $b\M$ is the compressive component of the Mach number \citep{Federrath2008,Konstandin2012a}, which is the Mach number associated with the compressive modes in the flow. It follows from Equation~\ref{eq:DenVarShockJump} that
    \begin{align} \label{eq:sigmaPar}
        \sigma^2_{\rho/\rho_0 \parallel} \approx b^2 \M^2.
    \end{align}

    \subsubsection{Type II: fast magnetosonic shocks}\label{sec:MHDplanarshock}

    Fast magnetosonic waves are the only MHD waves (in linearised MHD theory) that can propagate perpendicular to the magnetic field and compress the gas \citep{Landau1959,Lehmann2016,Tritsis2016}. The perpendicular waves propagate with velocity
    \begin{align} \label{eq:fastVelocity}
        v = \sqrt{v_{\rm A0}^2 + c_s^2},
    \end{align}
    where $v_{\rm A0} = B_0 / \sqrt{4\pi\rho}$ is the mean-field Alfv\'en velocity, and $c_s$ is the sound speed. The non-linear counterpart is the fast magnetosonic shock, which too propagates orthogonal to $\vecB{B}_0$. We call this a type~II shock. These shocks form through longitudinal compressions of the field lines, visualised in Figure~2 of \citet{Beattie2020c}. The field lines compress together, causing longitudinal, striated compressions in the density field. Through the flux-freezing condition, the magnetic field is locked perpendicular to the shock with $B \propto \rho$, hence these shocks are equivalent to the $B \propto \rho$ shocks described in \citet{Molina2012}. In an isothermal fluid the gas pressure scales directly with $\rho$ and hence the magnetic field becomes larger at higher $\rho/\rho_0$ in the fluid \citep{Mocz2018}. The shock jump is given by
    \begin{align} \label{eq:mhdJump}
        \left(\frac{\rho}{\rho_0}\right)_{\perp} = \dfrac{1}{2}\left[\sqrt{\left(1 + 2\frac{\MaO{2}}{b^2\M^2}\right)^2 + 8\MaO{2}} -\left(1 + 2\frac{\MaO{2}}{b^2\M^2}\right) \right],
    \end{align}
    and hence,
    \begin{align} \label{eq:sigmaPerp}
        \sigma^2_{\rho/\rho_0 \perp} \approx \dfrac{1}{2}\left[\sqrt{\left(1 + 2\frac{\MaO{2}}{b^2\M^2}\right)^2 + 8\MaO{2}} -\left(1 + 2\frac{\MaO{2}}{b^2\M^2}\right) \right].
    \end{align}
    Unlike the hydrodynamical shock-jump condition in Equation~\ref{eq:hdJump}, the density jump for fast magnetosonic waves is asymptotic to $(\rho/\rho_0)_{\perp} = (\sqrt{1 + 8\MaO{2}}-1)/2$ as $\M \rightarrow \infty$.  The limit is controlled entirely by the strength of the magnetic field. Physically, this corresponds to the strong magnetic field suppressing small-scale fluctuations that are introduced to the density as $\M$ increases (i.e., steepening of the $\rho$ power spectra; \citealt{Kim2005,Beattie2020}). If $\Mao \rightarrow 0$ then $(\rho/\rho_0)_{\perp} \rightarrow 0$. This is because the divergence of the velocity field only has a parallel to $\vecB{B}_0$ component as $B_0 \gg \delta B$, where $\delta B$ is the fluctuating component of the magnetic field. Hence one cannot create any perpendicular density contrasts for $B_0\to\infty$ \citep{Beattie2020c}.

    \subsection{Two-shock density variance model}\label{sec:twoShockModel}

    Now we combine Equation~\ref{eq:sigmaPar} and Equation~\ref{eq:sigmaPerp} to model the total variance of the anisotropic density field, i.e., we assume the stochastic medium of interest is composed out of the type~I and type~II shocks discussed above. Thus, we model the total fluctuations as the sum of integrals over the shock-jump conditions in the parallel and perpendicular directions, as discussed in \S\ref{sec:shockGeometry}, assuming that the fluctuations are independent from one another. In reality there will be some non-linear mixing of the shocks (e.g., type~II shocks compressing the type~I shocks longitudinally, changing the shock jump conditions through a multiplicative interaction) and, in principle, a larger diversity of shocks and density fluctuations, but we adopt the most parsimonious anisotropic model for the variance that one can formulate and leave generalisations of the model for future studies. The $N$-shock model would sum over $N$ types of \textit{uncorrelated} shocks with shock-jump conditions $\left(\rho/\rho_0\right)_{i}$ and shock volumes $\V_i$, given by
    \begin{equation}
        \sigma^2_{\rho/\rho_0} = \frac{1}{\V} \sum_{i}^{N} \int_{\V_i} \d{\V_i} \, \left[ \left(\frac{\rho}{\rho_0}\right)_{i} - 1 \right]^2.
    \end{equation}
    
    For our two-shock model ($N=2$, for type~I and type~II), this leads to

    \begin{widetext}
    \begin{align}
    \sigma_{\rho/\rho_0}^2 
    = & \sigma_{\rho/\rho_0\parallel}^2 + \sigma_{\rho/\rho_0\perp}^2=
    \underbrace{f_{\parallel} \int_{1}^{\rho / \rho_0} \d{\left(\frac{\rho_1}{\rho_0}\right)_{\parallel}} \,\left( \frac{\rho_1}{\rho_0} \right)^{-2}_{\parallel}  \left( \frac{\rho_1}{\rho_0} - 1 \right)^2_{\parallel}}_{\sigma^2_{\parallel}}
    +
    \underbrace{f_{\perp} \int_{1}^{\rho / \rho_0} \d{\left(\frac{\rho_1}{\rho_0}\right)}_{\perp} \,\left( \frac{\rho_1}{\rho_0} \right)_{\perp}^{-2} \left( \frac{\rho_1}{\rho_0} - 1 \right)^2_{\perp}}_{\sigma^2_{\perp}}, \label{eq:2shockModel_0}\\
    \sigma_{\rho/\rho_0}^2 = &  \underbrace{ f_{\parallel} b^2 \M ^2}_{\sigma^2_{\parallel}} +  \underbrace{ \frac{f_{\perp}}{2}\left[ \sqrt{\left(1 + 2\frac{\MaO{2}}{b^2\M^2}\right)^2 + 8\MaO{2}} -\left(1 + 2\frac{\MaO{2}}{b^2\M^2}\right) \right]}_{\sigma^2_{\perp}},\label{eq:2shockModel}
    \end{align}
    \end{widetext}
    \noindent where the parameters $f_{\parallel}$ and $f_{\perp}$ control the weighting of the contributions from each of the shock types, and come directly from integrating Equation~\ref{eq:densVariance} over two different sub-volumes that the fluctuations occupy. We call these the volume-fractions of the parallel and perpendicular fluctuations, respectively, and discuss and determine $f_{\parallel}$ and $f_{\perp}$ in \S\ref{sec:volumeFrac} below.

    \begin{figure*}
        \centering
        \includegraphics[width=1.0\linewidth]{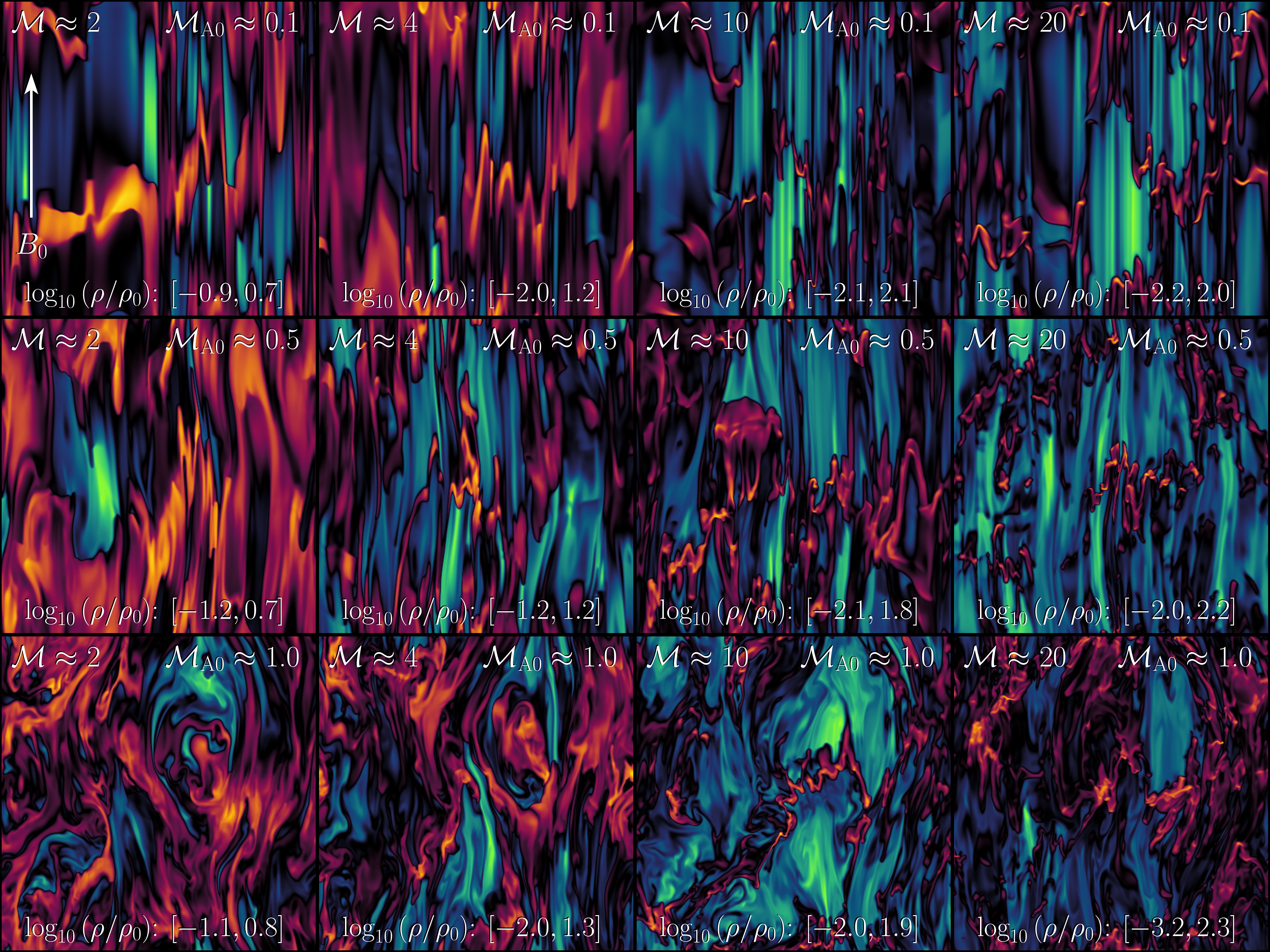}
        \caption{Slices through the $y=0$ plane of $\log_{10}(\rho/\rho_0)$ at $t = 6\,T$ for the 12 simulations listed in Table~\mbox{\ref{tb:simtab}}. The mean magnetic field, $\vecB{B}_0$, is oriented up the page as shown in the top-left panel. The plots are ordered such that the simulations with the highest Mach numbers are on the right column ($\M \approx 20$) and weakest ($\M \approx 2$) on the left. Likewise, the strongest $B_0$ simulations $(\Mao \approx 0.1)$ are on the top row and weakest ($\Mao \approx 1.0$) on the bottom. The top row reveals slices of the density structures for the most anisotropic simulations. For these simulations, fast magnetosonic waves (type~II shocks, \mbox{\S\ref{sec:MHDplanarshock}}) ripple through the density field, propagating perpendicular to $\vecB{B}_0$ and leaving striations from the compressions ($\rho/\rho_0 > 1$, shown in orange) and rarefactions ($\rho/\rho_0 < 1$, shown in green) in the field. The largest over-densities however form along the $\vecB{B}_0$, which are space-filling for low-$\M$, and narrow and highly-compressed for high-$\M$ (type~I shocks, \mbox{\S\ref{sec:HDplanarshock}}). For the $\Mao = 1.0$ simulations the anisotropy is beginning to weaken as the fluctuating component of the magnetic field grows, and becomes mixed into the fluid through the turbulence.}
        \label{fig:12_panel}
    \end{figure*}

    \begin{figure*}
        \centering
        \includegraphics[width=1.0\linewidth]{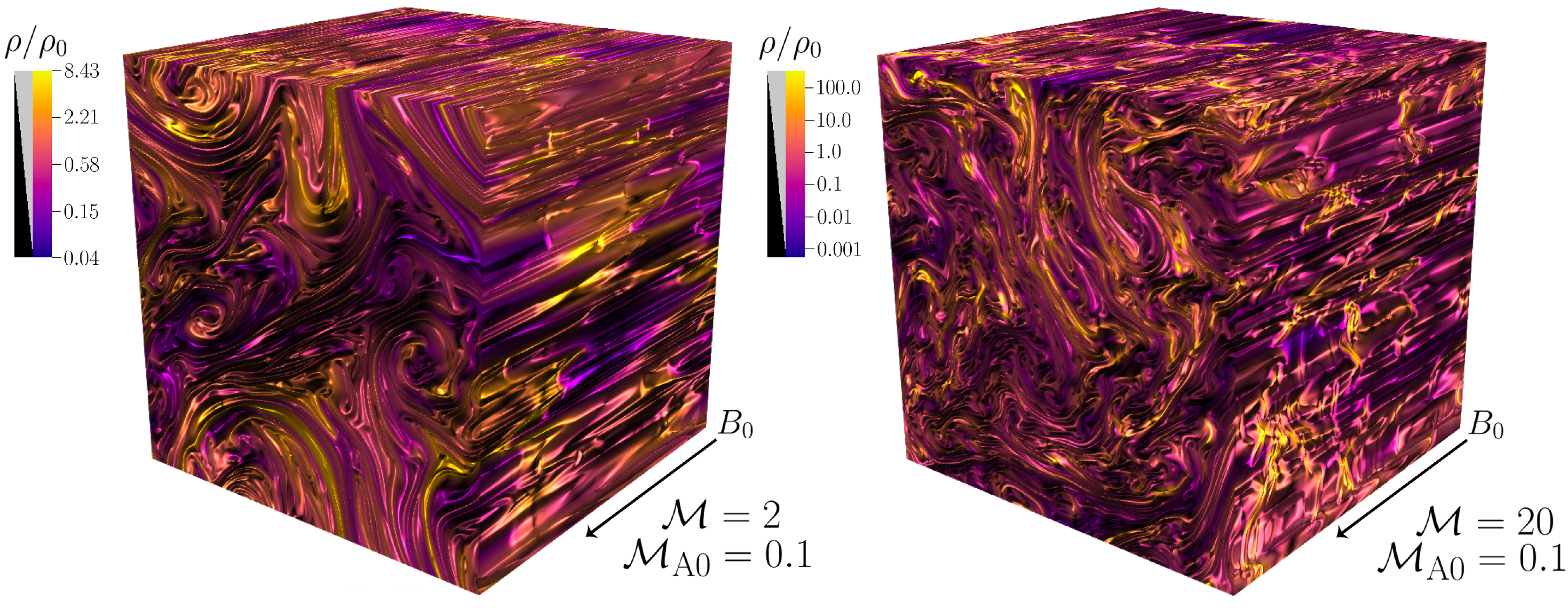}
        \caption{The structure of the 3D density field, $\rho/\rho_0$, for the highly-magnetised \texttt{M2MA0.1} (left) and \texttt{M20MA0.1} (right) simulations (see Table~\ref{tb:simtab}). The direction of the mean magnetic field, $\vecB{B}_0$, is shown at the bottom-right corner of the simulations. Large-scale vortices form in the density plane perpendicular to $\vecB{B}_0$. Density structures are tightly wound and stretched into ribbons and tubes as the supersonic vortices advect the flux-frozen density field around the field lines. The vortices are found on similar scales for both simulations, roughly $L/2$, the driving scale of the turbulence, however the \texttt{M20MA0.1} simulation has more small-scale density fluctuations than the \texttt{M2MA0.1} simulation (note the different scaling of the colour bars).}
        \label{fig:3dPlots}
    \end{figure*} 

    \section{MHD simulations}\label{sec:numerics}

    \subsection{MHD Model}

    Here we describe the numerical data that we use to test our density variance model. These are high-resolution, three-dimensional, ideal magnetohydrodynamical (MHD) simulations of supersonic turbulence. The ideal, isothermal MHD equations are
    \begin{align}
        \frac{\partial \rho}{\partial t} + \nabla\cdot(\rho \vecB{v}) &= 0 \label{eq:continuity}, \\
        \rho \left( \frac{\partial}{\partial t} + \vecB{v}\cdot\nabla \right) \vecB{v} &= \frac{(\vecB{B} \cdot \nabla)\vecB{B}}{4\pi} - \nabla \left(c_s^2 \rho + \frac{|\vecB{B}|^2}{8\pi}\right) + \rho \vecB{F},\label{eq:momentum} \\ 
        \frac{\partial \vecB{B}}{\partial t} &= \nabla \times (\vecB{v} \times \vecB{B}), \label{eq:induction} \\
        \nabla \cdot \vecB{B} &= 0, \label{eq:div0}
    \end{align}
    where $\vecB{v}$ is the fluid velocity, $\rho$ the density, $\vecB{B}$ the magnetic field, $c_s$ the sound speed and $\vecB{F}$, a stochastic function that drives the turbulence. We use a modified version of \textsc{flash} based on version~4.0.1 \citep{Fryxell2000,Dubey2008} to solve the MHD equations in a periodic box with dimensions $L^3$, on a uniform grid with resolution $512^3$, using the multi-wave, approximate Riemann solver framework described in \citet{Bouchut2010} and implemented in \textsc{flash} by \citet{Waagan2011}. For a detailed discussion of the simulations we refer the reader to \citet{Beattie2020} and \citet{Beattie2020c}. Table~\ref{tb:simtab} provides a summary of the important time-averaged parameter values. Here we briefly summarise the key methods and properties of the simulations.
    
   \begin{table}
        \caption{Main simulation parameters.}
        \centering
        \begin{tabular}{l r@{}l r@{}l r@{}l l}
            \hline
            \hline
            Sim. ID & \multicolumn{2}{c}{$\M\,(\pm1\sigma)$} & \multicolumn{2}{c}{$\Mao\,(\pm1\sigma)$} & \multicolumn{2}{c}{$\sigma_{\rho/\rho_0}^2\,(\pm1\sigma)$}  & $N^3$ \\
            (1) & \multicolumn{2}{c}{(2)} & \multicolumn{2}{c}{(3)} & \multicolumn{2}{c}{(4)} & (5) \\
            \hline 
            \texttt{M2Ma0.1}       & 2.6\,    & $ \pm$ 0.2      & 0.131\, & $ \pm$ 0.008 &  0.49 \,     & $ \pm$ 0.05    & $512^3$   \\
            \texttt{M4Ma0.1}       & 5.2\,    & $ \pm$ 0.4      & 0.13\,  & $ \pm$ 0.01  &  1.6\,     & $ \pm$ 0.3    & $512^3$   \\
            \texttt{M10Ma0.1}      & 12\,     & $ \pm$ 1        & 0.125\, & $ \pm$ 0.006 &  5.3\,     & $ \pm$ 0.9     & $512^3$   \\
            \texttt{M20Ma0.1}      & 24\,     & $ \pm$ 1        & 0.119\, & $ \pm$ 0.003 &  7\,     & $ \pm$ 1      & $512^3$   \\
            \texttt{M2Ma0.5}       & 2.2\,    & $ \pm$ 0.2      & 0.54\, & $ \pm$ 0.04   &  0.51\,     & $ \pm$ 0.07    & $512^3$   \\
            \texttt{M4Ma0.5}       & 4.4 \,   & $ \pm$ 0.2      & 0.54\, & $ \pm$ 0.03   &  1.8\,     & $ \pm$ 0.3    & $512^3$   \\
            \texttt{M10Ma0.5}      & 10.5\,   & $ \pm$ 0.5      & 0.52\, & $ \pm$ 0.02   &  5.1\,     & $ \pm$ 0.7    & $512^3$   \\
            \texttt{M20Ma0.5}      & 21 \,    & $ \pm$ 1        & 0.53\, & $ \pm$ 0.02   &  8\,     & $ \pm$ 1      & $512^3$   \\
            \texttt{M2Ma1.0}       & 2.0\,    & $ \pm$ 0.1      & 0.98\, & $ \pm$ 0.07   &  0.5\,     & $ \pm$ 0.1    & $512^3$   \\
            \texttt{M4Ma1.0}       & 3.8\,    & $ \pm$ 0.3      & 0.95\, & $ \pm$ 0.08   &  2.0\,     & $ \pm$ 0.3    & $512^3$   \\
            \texttt{M10Ma1.0}      & 9.3\,    & $ \pm$ 0.5      & 0.93\, & $ \pm$ 0.05   &  8\,     & $ \pm$ 1    & $512^3$   \\
            \texttt{M20Ma1.0}      & 18.8\,   & $ \pm$ 0.7      & 0.93\, & $ \pm$ 0.03   &  12\,     & $ \pm$ 2    & $512^3$   \\
            \hline 
            \hline
        \end{tabular} \\
        \begin{tablenotes}
        \item{\textit{\textbf{Notes:}} For each simulation we extract 51~realisations at times $t/T\in\left\{5.0,\,5.1,\,\hdots,\,10.0\right\}$, where $T$ is the turbulent turnover time, and all $1\sigma$ fluctuations listed are from the time-averaging over the $5\, T$ time span. Column (1): the simulation ID. Column (2): the rms turbulent Mach number, $\M = \sigma_V / c_s$. Column (3): the Alfv\'en Mach number for the mean-$\vecB{B}$ component, $\vecB{B}_0$, $\Mao = (2c_s\M\sqrt{\pi \rho_0}) / |\vecB{B}_0|$, where $\rho_0$ is the mean density, $c_s$ is the sound speed. Column (4) the volume-weighted density variance, computed as shown in Equation \ref{eq:densVariance}. Column (5): the number of computational cells for the discretisation of the spatial domain of size $L^3$.}
        \end{tablenotes}
        \label{tb:simtab}
    \end{table}

    \subsection{Turbulent driving, density and velocity fields}

    The initial velocity field is set to $\vecB{v}(x,y,z,t=0)=\vecB{0}$, with units $c_s=1$, and the density field $\rho(x,y,z,t=0)=\rho_0$, with units $\rho_0=1$. The turbulent acceleration field, $\vecB{F}$, in Equation~\ref{eq:momentum} follows an Ornstein-Uhlenbeck process in time and is constructed such that we can control the mixture of solenoidal and compressive modes in $\vecB{F}$ (see \citealt{Federrath2008,Federrath2009,Federrath2010} for a detailed discussion of the turbulence driving). We choose to drive with a natural mixture of the two modes \citep{Federrath2010}, which corresponds to $b \approx 0.4$. We isotropically drive in wavenumber space, centred on $k=2$, corresponding to real-space scales of $\ell_D = L/2$, and falling off to zero with a parabolic spectrum between $k=1$ and $k=3$. Thus, energy is injected only on large scales and the turbulence on smaller scales develops self-consistently through the MHD equations. The auto-correlation timescale of $\vecB{F}$ is equal to 
    \begin{align} \label{eq:T(M)}
    T = \frac{\ell_D}{\sigma_V} = \frac{L}{2 c_s \M}.
    \end{align}
    We vary the sonic Mach number between $\M = 2$ and 20, encapsulating the range of observed $\M$ values for turbulent MCs (e.g., \citealt{Schneider2013,Federrath2016b,Orkisz2017,Beattie2019b}). We run the simulations from $0 \leq t/T \leq 10$ and extract 51~time realisations of $\rho(x,y,z)$ over $5 \leq t/T \leq 10$ to gather data only when the turbulence is in a statistically stationary state \citep{Federrath2009,Price2010}. For simulations with a strong guide field, a statistically steady state is reached after about $5T$, while for purely hydrodynamical turbulence and super-Afv\'enic turbulence, stationarity is already reached after about $2T$. All our results are based on time-averages across these 51~realisations, unless explicitly indicated otherwise.
    
    \begin{figure}
        \centering
        \includegraphics[width=\linewidth]{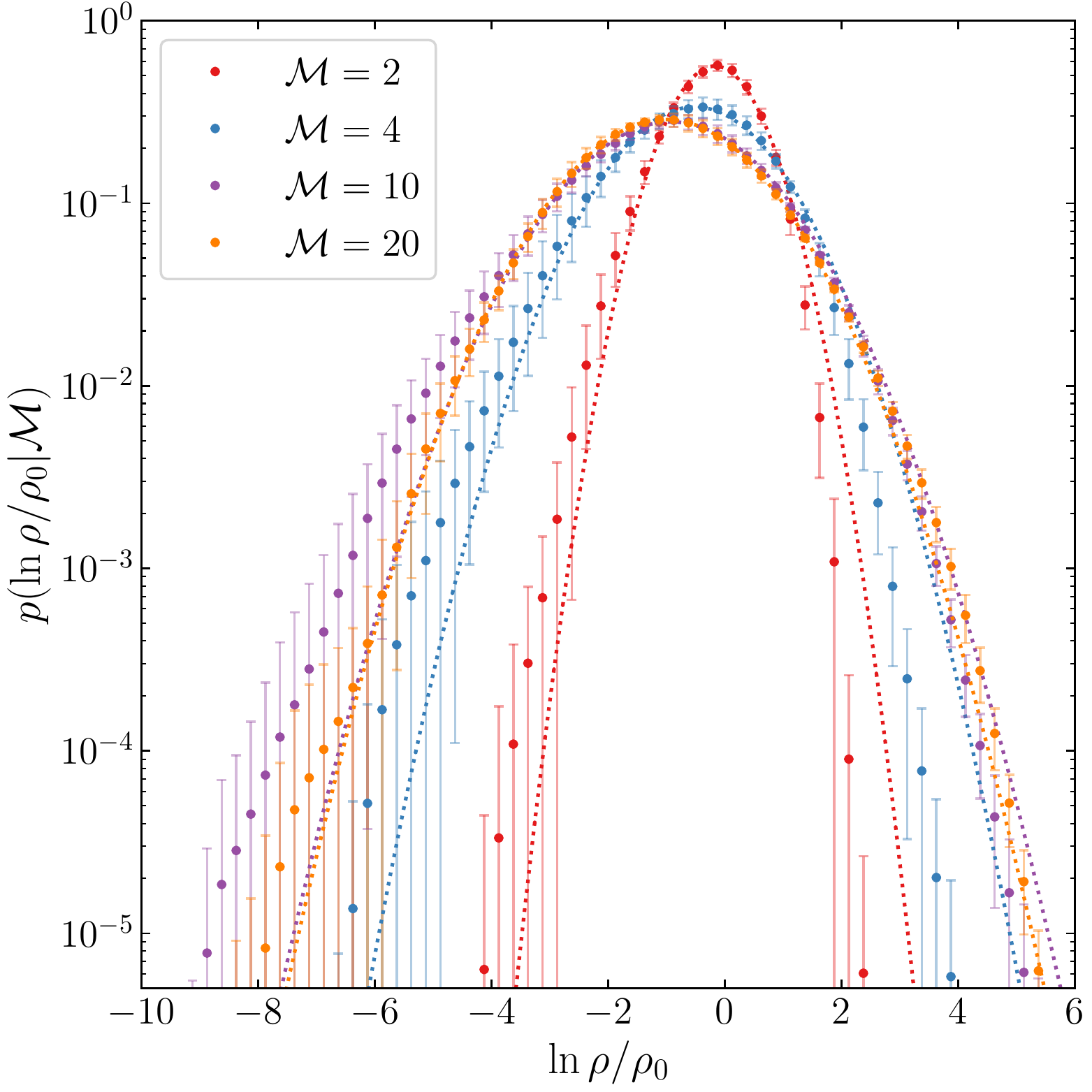}
        \caption{Time-averaged logarithmic density PDFs for the ensemble of simulations, $2 \lesssim \M \lesssim 20$ with $\Mao = 0.1$. We show log-normal curves overlaid with dotted lines on the data. Consistent with findings in \mbox{\citet{Molina2012}}, for sub-Alfv\'enic mean-field, compressible turbulence we find that the density fluctuations are approximately log-normal, and do not exhibit strong skewness features that are present in hydrodynamical turbulence \mbox{\citep{Federrath2010,Hopkins2013,Mocz2019}}, and hence are well described by the variance of the density field.}
        \label{fig:dens_pdf}
    \end{figure}
    
    In Figure~\mbox{\ref{fig:dens_pdf}} we show the time-averaged logarithmic density PDFs for the $\Mao = 0.1$ simulations shown in Table~\mbox{\ref{tb:simtab}}. We show a log-normal model fit to the density fluctuations with dotted lines. The density PDFs show that the fluctuations are well described by a log-normal model, especially as $\M$ increases. This is in contrast with hydrodynamical and super-Alfv\'enic compressible turbulent flows, which become less log-normal as the $\M$ increases \citep{Federrath2010,Hopkins2013,Mocz2019}. We leave the detailed analysis of the Gaussianity of the density PDFs for a follow-up study, but stress that in the sub-Alfv\'enic regime the variance of the density field, which is the focus of this study, is an informative statistic because of the log-normal fluctuations.
    
    In Figure~\mbox{\ref{fig:12_panel}} we show density slices through a plane perpendicular to the direction of $\vecB{B}_0$, indicated in the top-left plot. The plots are organised such that the top-left plot has the weakest (but still supersonic) turbulence ($\M = 2$) and strongest $B_0$ ($\Mao = 0.1$), and the bottom-right plot has the strongest turbulence ($\M = 20$) and weakest $B_0$ ($\Mao = 1.0$). The slices reveal the highly-anisotropic density structures that form in sub-Alfv\'enic, supersonic turbulence. Over-densities and rarefactions from fast magnetosonic shocks (type~II shocks) cause ripples through the density running parallel to the magnetic field. The largest density contrasts are from the shocks that form along the magnetic field (type~I shocks), which compress the gas density into sharp, fractal filaments. The filamentary structures become less space-filling as $\M$ increases, extending across $\propto L/\M^2 \propto L/4$ for $\M \approx 2$, to a tiny fraction of the box at $\M \approx 20$, qualitatively consistent with the shock-width model discussed in \S\mbox{\ref{sec:shockGeometry}}.
    
    Using the \textsc{ospray} ray-tracer in \textsc{visit} \citep{Childs2012} we show examples of the volumetric density field for the \texttt{M2MA0.1} and \texttt{M20MA0.1} simulations in Figure~\ref{fig:3dPlots}. Large-scale vortices are revealed in the \texttt{M2MA0.1} simulation, with sheets of density wrapped around in the direction of the $\vecB{B}_0$. The vortices form at roughly the driving scale ($L/2$) of the simulation. The volumetric rendering for \texttt{M20MA0.1} shows much more small-scale structure compared to \texttt{M2MA0.1}, both along and perpendicular to the direction of $\vecB{B}_0$. The large-scale vortices found in \texttt{M2MA0.1} are roughly at the same scale as the vortices in \texttt{M20MA0.1}, suggesting that it is the driving scale that sets the size of the vortices. Dense, shocked material is compressed along the magnetic field lines, creating the filamentary structures that were found in the slices, oriented perpendicular to the field.

    \subsection{Magnetic fields} \label{sec:Bfields}

    The initial magnetic field, $\vecB{B}(x,y,z) = B_0\hat{\vecB{z}}$ at $t=0$, in Equations~(\mbox{\ref{eq:momentum}--\ref{eq:div0}}) is a uniform field with field lines threaded through the $\hat{\vecB{z}}$ direction of the simulations. $B_0$ is set to ensure the desired $\Mao$, using the definition of the Alfv\'en velocity (see footnote \ref{foot:Ma0}) and the turbulent Mach number (see Equation~\ref{eq:M}),
    \begin{equation}\label{eq:meanField}
    \Mao = \sigma_V / V_{\rm A0} = 2c_s\sqrt{\pi\rho_0}\M/B_0,
    \end{equation}
    with $B_0$ constant in space and time. We set $\Mao = 0.1, 0.5$ and $1.0$ for different simulations (see Table~\ref{tb:simtab}), ensuring that the turbulence is in an anisotropic regime \citep{Beattie2020c,Beattie2020}. The total magnetic field is given by
    \begin{align}
    \vecB{B}(t) &= B_0 \vecB{\hat{z}} + \delta\vecB{B}(t), 
    \end{align}
     where the fluctuating component of the field, $\delta \vecB{B}$, evolves self-consistently from the MHD equations, with $\Exp{\delta\vecB{B}}_t = 0$. From previous experiments in the anisotropic regime discussed throughout this study, $|\delta\vecB{B}|/|\vecB{B}_0| \approx 10^{-3} -10^{-2} $ \citep{Federrath2016c,Beattie2020c}. For this reason $\Ma \approx \Mao$ in this regime, however, we explicitly formulate our models around $\Mao$, which is an invariant across different $\M$ simulations. 

    \section{Implementation of the anisotropic density variance model}\label{sec:varianceModel}
    In this section we describe and model the fluctuation volume-filling fractions that we introduced in Equation~\ref{eq:2shockModel}. 
    
    \subsection{Volume fractions of anisotropic fluctuations}\label{sec:volumeFrac}
    The fluctuation volume fractions, $f_{\parallel}$ and $f_{\perp}$ in Equation~\mbox{\ref{eq:2shockModel}}, determine the contributions of the parallel and perpendicular fluctuations in our model. These parameters naturally come about from considering multiple terms of a volume-weighted statistic, where the total volume is partitioned into distinct sub-volumes for the parallel and perpendicular fluctuations. The volume-filling fractions can be derived by returning to Equation~\mbox{\ref{eq:shockVol}}, but instead of using a planar shock, we consider shocks that have a reduced volume-filling factor, $f_i$,
    \begin{align} \label{eq:shockVolD}
        \V_{\rm shock, i} = f_{i}\lambda L^{2},
    \end{align}
    where $i \in \left\{\parallel , \perp\right\}$ is the volume-filling factor of the shock, e.g., for $f_i = 1$ the shock is the planar shock from Equation~\mbox{\ref{eq:shockVol}}, and for $f_i < 1$ the shock has a smaller volume than a planar shock, which, could be for example, a tubular, filamentary shock. Considering shocks with a variable volume is an important consideration to make in our two-shock model, because the amount of volume that the fluctuations fill encodes how much they contribute to the variance through the density variance integral, Equation~\mbox{\ref{eq:densVariance}}. Propagating Equation~\mbox{\ref{eq:shockVolD}} through Equations~\mbox{\ref{eq:dVdRho}}-\mbox{\ref{eq:DenVarShockJump}}, making the same assumptions but for non-planar shocks, shows that
    \begin{align}
        \sigma_{\rho/\rho_0}^2 \approx & f_{\parallel} \left[ \left(\frac{\rho}{\rho_0}\right)_{\parallel} - \left(\frac{\rho_0}{\rho}\right)_{\parallel} - 2\ln\left( \frac{\rho}{\rho_0} \right)_{\parallel}\right] \\
        &+ f_{\perp}\left[\left(\frac{\rho}{\rho_0}\right)_{\perp} - \left(\frac{\rho_0}{\rho}\right)_{\perp} - 2\ln\left( \frac{\rho}{\rho_0} \right)_{\perp}\right],
    \end{align}
    where $f_{\parallel}$ is the volume-filling fraction of the type~I shocks and $f_{\perp}$ is the volume fraction of the type~II shocks. Note that the geometry of the volume variable shock propagates through to the total volume fraction of the parallel and perpendicular fluctuations, including the rarefactions. This is because the rarefaction region in the flow will share some of the same geometrical properties as the shock, since the gas density is compressed from the rarefaction into the over-density. However, as we discussed in \S\mbox{\ref{sec:shockGeometry}}, the main contributors to the variance are the over-densities, hence,
    \begin{align}
        \sigma_{\rho/\rho_0}^2 \approx & f_{\parallel} \left(\frac{\rho}{\rho_0}\right)_{\parallel} + f_{\perp} \left(\frac{\rho}{\rho_0}\right)_{\perp},
    \end{align}
    and Equation~\mbox{\ref{eq:2shockModel}} immediately follows by substituting in the appropriate shock-jump conditions. Since the volume-filling fractions are for the total parallel and perpendicular fluctuations we assume that they add to unity,
    \begin{align} \label{eq:volumeCriteria}
        1 = \sum_{i}^{N} f_i \iff 1 = f_{\parallel} + f_{\perp}.
    \end{align}
    This means our model is an $N-1$ parameter model, where $N$ is the number of shock types, and more explicitly, a one-parameter model for the two-shock model we describe here.
    
    The volume-filling fractions need not be constants, and indeed, may depend upon both $\M$ and $\Mao$. For example, \citet{Konstandin2016}, \citet{Beattie2019a} and \citet{Beattie2019b} demonstrate that the global fractal dimension $\D$, which is a measure of the how the most singular structures in the flow fill space, for supersonic turbulence depends upon $\M$, which varies between 3 (space-filling, low-$\M$) and 1 (filaments and tubular shocks, high-$\M$). This is because the flow is more compressible for higher $\M$, reducing $\D$, which corresponds to the emergence of highly-compressed structures like one-dimensional filaments. Increasing $\M$ also directly affects the geometry of the shocks in the flow, which can be shown by expressing Equation~\ref{eq:shockThick}, the shock thickness, in terms of the shock-jump conditions for an isothermal, hydrodynamical shocks in pressure equilibrium with the ambient environment,
    \begin{align}
         \lambda \approx \frac{L}{\M^2}.
    \end{align}
    This establishes that the shocks become more compressed (thinner), filling less space, as $\M$ increases, which means that a reasonable model for the volume fraction of the parallel fluctuations, which is dominated by type I shocks is $f_{\parallel}\propto\M^{-2}$. However, this only holds for $\M\gg1$, but not for $\M\sim1$.  Thus, we require a phenomenological function for the volume fraction that describes its dependence on $\M$ for both the trans to supersonic flow regime. We consider the form
    \begin{equation} \label{eq:vfrac}
        f_{\parallel} = f_0\left[1 + \left(\frac{\M}{\M_{\rm c}}\right)^4 \right]^{-1/2},
    \end{equation}
    which describes a function that tends towards $\sim 1/\M^2$ when $\M > \M_{\rm c} \gg 1$ and a constant volume fraction, $f_0$, for $\M < \M_{\rm c}$, e.g., in the sub- trans-sonic regime, where the type I shock thickness becomes ill-defined, and linear MHD waves (slow, parallel and fast, perpendicular modes) weakly compress the flow. Here, $\M_{\rm c}$ defines a critical Mach number for the transition between the constant and the $\sim \M^{-2}$ regime. Thus, the parameter $f_0$ defines the volume-filling fraction of the parallel fluctuations, in the presence of over-densities formed by slow modes, as discussed in \S\ref{sec:HDplanarshock}. These are acoustic modes that propagate along the magnetic field in the subsonic regime \citep{Landau1959}. Note that since we have assumed $f_{\perp} = 1 - f_{\parallel}$ we only need to consider a model for $f_{\parallel}$.

    \begin{figure}
        \centering
        \includegraphics[width=\linewidth]{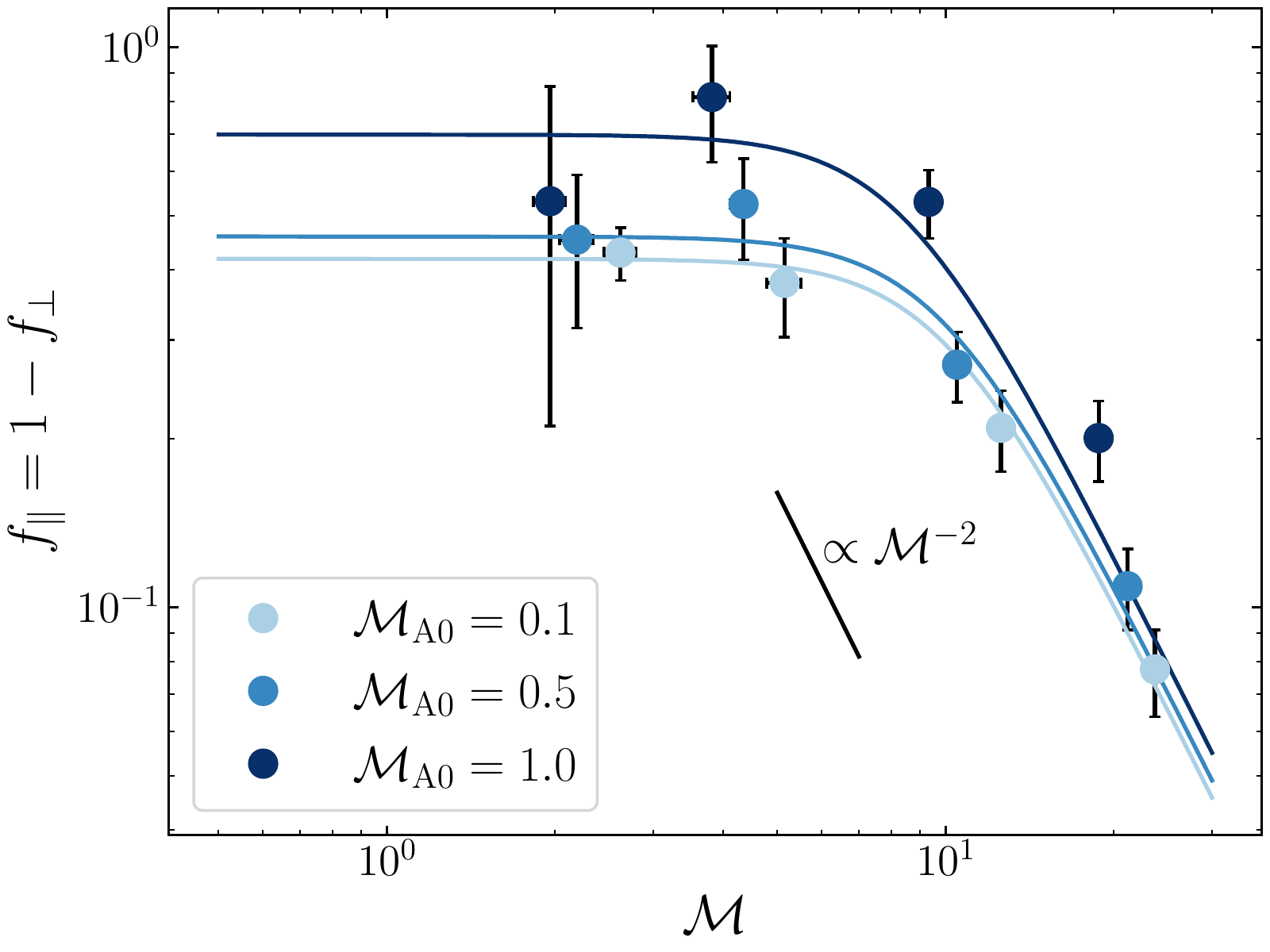}
        \caption{The volume-filling fraction of fluctuations that run parallel to $\vecB{B}_0$, $f_{\parallel}$, as a function of $\M$, as discussed in \S\ref{sec:volumeFrac}, for all of the trans/sub-Alfv\'enic simulations from Table~\ref{tb:simtab}, with $1\sigma$ uncertainties propagated from Equation~\ref{eq:fpar_analytic}. We fit our phenomenological models in Equation~\ref{eq:vfrac} to each of the $\Mao$ datasets, shown the with solid lines. This models suggests that $f_{\parallel} \sim \lambda \sim \M^{-2}$ in the high-$\M$ limit, where $\lambda$ is the shock-width, and $f_{\parallel} \sim \text{const.}$ in the low-$\M$ limit, which is supported well by the data. This means, assuming $f_{\perp} = 1 - f_{\parallel}$, $f_{\perp} \gg f_{\parallel}$ for high-$\M$ flows, i.e., the perpendicular fluctuations, which are dominated by fast magnetosonic shocks, contribute most to volume fraction at high-$\M$, which is visualised in Figure~\ref{fig:divDecomp}.}
        \label{fig:volumeFrac}
    \end{figure}

    \begin{figure*}
        \centering
        \includegraphics[width=\linewidth]{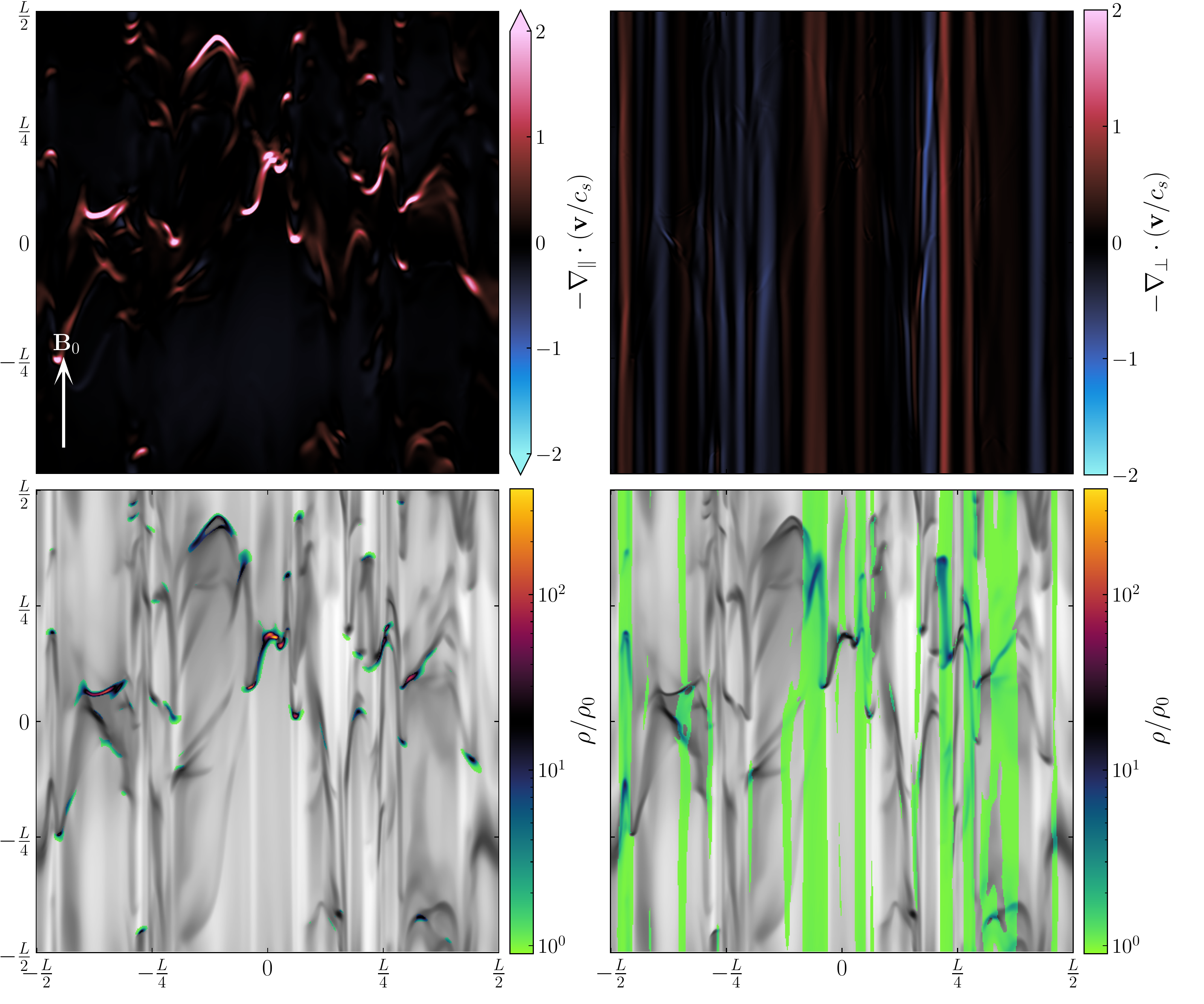}
        \caption{\textbf{Top left:} A slice of the parallel convergence of the velocity field, with respect to the direction parallel to $\vecB{B}_0$. The convergence is in units of $c_s/L$. Here we show a single time realisation from the $\M = 20$, $\Mao = 0.1$ simulation (listed in Table~\ref{tb:simtab}), revealing hydrodynamical shocks that propagate parallel to $\vecB{B}_0$, reminiscent in morphology of the hydrodynamical bow shocks studied in \citet{Robertson2018}. \textbf{Top right:} The same time realisation as the left plot, but for the perpendicular convergence of the velocity field, showing fast magnetosonic compressions of the velocity field, propagating orthogonal to $\vecB{B}_0$. The amplitude of the convergence is an order of magnitude larger for the hydrodynamical shocks compared to the magnetosonic fast shocks due to the strong magnetic cushioning effect perpendicular to $\vecB{B}_0$ \citep{Molina2012}. \textbf{Bottom left:} the parallel shocks in the density field, corresponding to the positive convergent structures in the upper-left plot, with density contrasts up to $(\rho/\rho_0)_{\parallel} \sim \M^2 = 400$, and very low volume-filling fractions. These are the type~I shocks discussed in \S\ref{sec:HDplanarshock}, which form the hydrodynamical component of our density variance model, shown as $\sigma_{\rho/\rho_0\parallel}^2$ in Equation~\ref{eq:2shockModel}. \textbf{Bottom right:} the same as the bottom left plot, but for the perpendicular shocks in the density field. The volume-filling fraction of the perpendicular fluctuations is much larger than the parallel component, however the density contrast is orders of magnitude lower. These are the type~II shocks outlined in \S\ref{sec:MHDplanarshock}, which form the magnetohydrodynamical component of our density variance model, $\sigma_{\rho/\rho_0\perp}^2$.}
        \label{fig:divDecomp}
    \end{figure*}
    
    \subsection{Determining the volume fraction of parallel and perpendicular fluctuations}\label{sec:f_par_data}
    The first step to fitting our model is to develop a dataset, ($\M_i$,$f_{\parallel,i}$), for the different $\Mao$ simulation listed in Table~\ref{tb:simtab}. Since, as discussed in \S~\ref{sec:volumeFrac}, our model only has a single parameter, $f_{\parallel}$, we can rearrange Equation~\ref{eq:2shockModel} and solve analytically\footnote{This follows from $\sigma^2_{\rho/\rho_0} = f_{\parallel}\sigma^2_{\rho/\rho_0\parallel} + f_{\perp}\sigma^2_{\rho/\rho_0\perp} = f_{\parallel}\sigma^2_{\rho/\rho_0\parallel} + (1 - f_{\parallel})\sigma^2_{\rho/\rho_0\perp}$ and then solving for $f_{\parallel}.$},
    \begin{equation}\label{eq:fpar_analytic}
        f_{\parallel} = \frac{\sigma^2_{\rho/\rho_0} - \sigma^2_{\rho/\rho_0\perp}}{\sigma^2_{\rho/\rho_0\parallel} - \sigma^2_{\rho/\rho_0\perp}},
    \end{equation}
    with the single constraint that $f_{\parallel} \in [0,1]$, since it corresponds to a fraction of the total volume. This allows us to generate a function $f_{\parallel}(\M)$ for each simulation. We show this data in Figure~\ref{fig:volumeFrac}. We then fit the functional form for $f_{\parallel}(\M)$, Equation~\ref{eq:vfrac}, with parameters $f_0$ and $\M_{\rm c}$ using a non-linear least squares fitting routine weighted by $1/\Delta f_{\parallel}^2$, where $\Delta f_{\parallel}$ is the uncertainty in $f_{\parallel}$, propagated through Equation~\ref{eq:fpar_analytic}. We show the fits using the solid lines, coloured by $\Mao$. 

    \begin{table}
        \caption{Fit parameters for $f_{\parallel}(\M)$, as per Equation \ref{eq:vfrac}.}
        \centering
        \begin{tabular}{c r@{}l r@{}l r@{}l}
        \hline
        \hline
        Parameter & \multicolumn{2}{c}{$\Mao=0.1$} & \multicolumn{2}{c}{$\Mao = 0.5$} & \multicolumn{2}{c}{$\Mao=1.0$} \\
        (1) & \multicolumn{2}{c}{(2)} & \multicolumn{2}{c}{(3)} & \multicolumn{2}{c}{(4)} \\
        \hline
        $f_0$ & 0.42\, & $ \pm$ 0.05 & 0.46\, & $ \pm$ 0.07 & 0.71\, & $ \pm$ 0.08 \\
        $\M_{\rm c}$ & 9.9\, & $ \pm$ 0.2 & 10\, & $ \pm$ 1 & 10\, & $ \pm$ 1 \\
        \hline
        \hline
        \end{tabular} \\
        \label{tb:parms}
    \end{table}

    For each of the datasets, the critical sonic Mach number is $\M_{\rm c} \approx 10$ and the $f_0$ parameter varies monotonically between $\approx 0.4$--$0.7$, listed in Table~\ref{tb:parms}. There are two important conclusions to make: (1) magnetised turbulence is significantly saturated with shocks above $\M \approx 10$, and (2) the parallel fluctuations become less confined by the magnetic field, and occupy more of the total volume when the magnetic field is weakened. Accordingly, one should expect that $f_0 \sim 1.0$ as $\Mao \rightarrow \infty$, i.e., when there is no confinement from the magnetic field. This reclaims the hydrodynamical $\sigma_{\rho/\rho_0}^2 - \M$ relation. Regardless of the field strength, the high-$\M$ behaviour is the same between the simulations, which is expected since the shock width, $\lambda \sim \M^{-2}$, encodes how the $(b\M)^2$ term contributes to the total variance once the fluid is sufficiently shocked. This transition is consistent with what \citet{Beattie2020} found, where $\M \approx 4 - 10$ marked the Mach number for when the anisotropy of the 2D power spectrum revealed a morphology dominated by shocks aligned perpendicular to $\vecB{B}_0$. Since we have assumed $f_{\perp} = 1 - f_{\parallel}$ (Equation~\ref{eq:volumeCriteria}) as $\M$ grows and the parallel fluctuations occupy less and less of the volume until the perpendicular fluctuations contribute the most to the total volume budget for the fluid.
    
    We show some of the shocked regions for both the parallel and perpendicular fluctuations in the top panels of Figure~\ref{fig:divDecomp}, for $\M \approx 20$ turbulence by taking the divergence along and across $\vecB{B}_0$. We also show the corresponding density fluctuations in the bottom two panels, of the Figure, where we highlight the $-\nabla\cdot\vecB{v} > 0$ regions. We find that the relative fraction of the volume occupied by type~I (left-hand panel) and type~II (right-hand panel) shocks is qualitatively consistent with our model, as demonstrated by the regions of high compression quantified by $-\nabla\cdot\vecB{v} > 0$ in the top panels, and corresponding density structures coloured in the bottom panels. This Figure is primarily for illustrative purposes of type~I and type~II shocks and their approximate volume filling fractions.
    
    \begin{figure*}
        \centering
        \includegraphics[width=\linewidth]{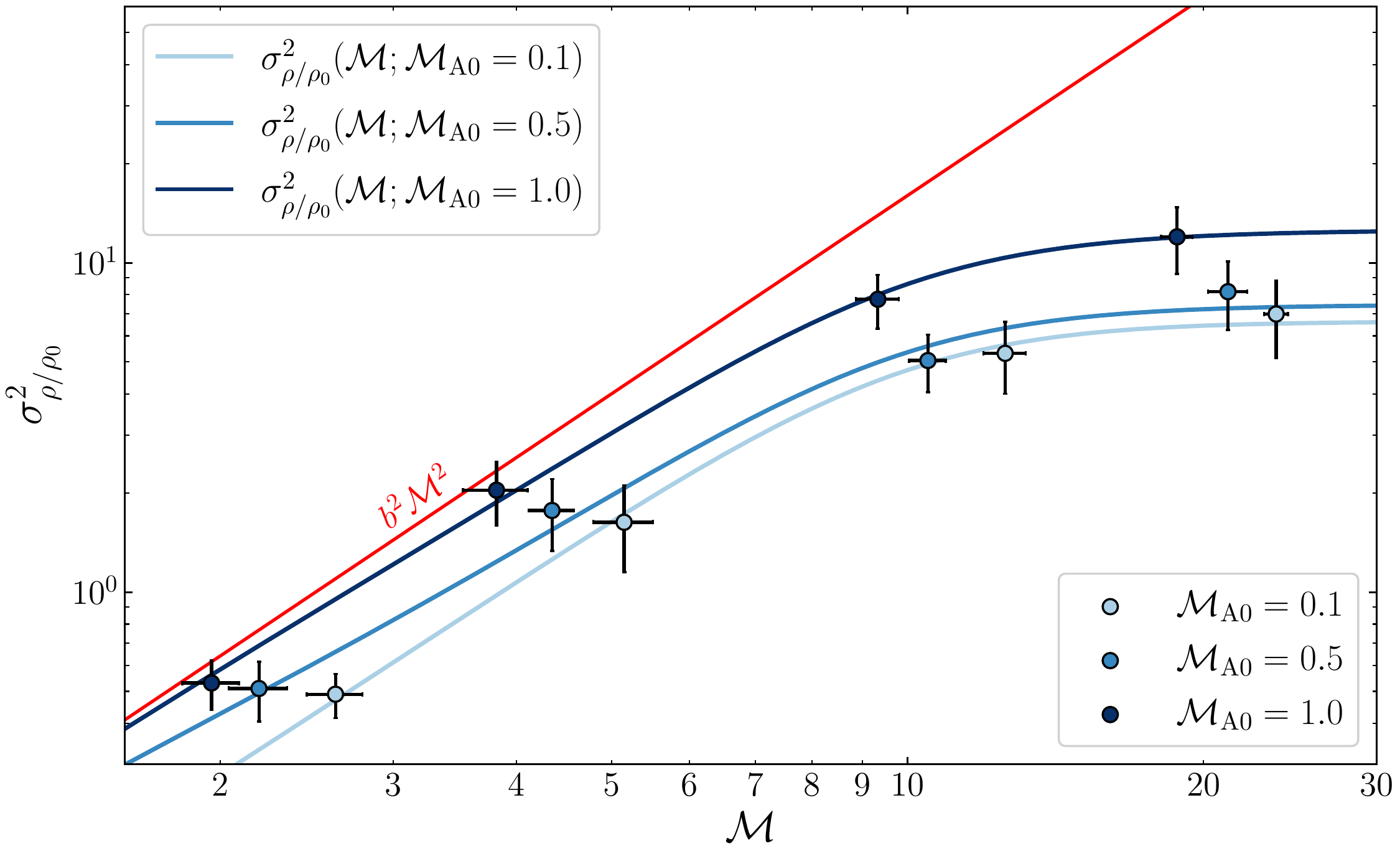}
        \caption{Two-shock model (Equation~\ref{eq:2shockModel}) for the density variance, $\sigma^2_{\rho/\rho_0}$, as a function of $\M$, for each set of $\Mao$ simulations. Blue points are numerical data from the 12~simulations (Table~\ref{tb:simtab}), averaged over a time interval of $5\,T$. The red line indicates the hydrodynamical limit for the density variance, $b^2\M^2$. We find that the variance grows from low $\M$ to $\M \approx 4$ in a hydrodynamical fashion, but then saturates as the small-scale density fluctuations (associated with high-$\M$ values) are suppressed by the strong mean magnetic field. We plot our model, Equation~\ref{eq:2shockModel}, on each $\Mao$ dataset, shown with the solid lines. We find good agreement between the data and our models, which suggests there is an upper bound for $\sigma^2_{\rho/\rho_0}$ in this highly-magnetised, anisotropic turbulent regime. We discuss the implications for this result in \S\ref{sec:discussion}.}
        \label{fig:varModelFit}
    \end{figure*} 
    
    \section{Anisotropic density variance model results}\label{sec:results}
    Now that we have constrained $f_{\parallel}$ we put this back into Equation~\ref{eq:finalModel} and generate our final variance model. We depict our models with solid lines in Figure~\ref{fig:varModelFit}. The hydrodynamical limit is drawn in red, which provides an upper bound of the variance ($f_{\parallel} = 1$, $f_{\perp} = 0$), and the three sets of simulation data at different $\Mao$ in blue. The models fit the data well, never deviating from the data by more than a small fraction of the total 1$\sigma$ fluctuations in $\sigma_{\rho/\rho_0}^2$.
    
    Our results establish the following picture of the density field in this regime: at low $\M$, hydrodynamical, parallel to $\vecB{B}_0$, fluctuations are large-scale, occupying a significant fraction of the fluid volume with a relatively large volume-filling fraction. The type~I shocks dominate the parallel fluctuation contribution to the volume-weighted variance of the density field. However, as $\M$ increases, the small-scale fluctuations grow in the field, with and the type~I shocks shrink in size with their shock width $\sim \M^{-2}$, and thus the parallel fluctuations begin to occupy only very small fractions of the fluid volume, decreasing the contribution from type~I shocks to the variance. When $\M \gg 1$, the total volume-weighted variance becomes  by type~II shocks, which are highly magnetised and do not support small-scale fluctuations. This flattens $\sigma_{\rho/\rho_0}^2(\M)$ out, because the small-scale fluctuations that are added to the field with increasing $\M$ are suppressed. Eventually this leads to an upper bound on $\sigma_{\rho/\rho_0}^2$, as the type~II shock density contrasts dominate the total variance.
    
    \subsection{Limiting behaviour of the model}
    Let us now consider the limiting behaviour of the density variance in our model. The density variance is summarised as
    \begin{align}\label{eq:finalModel}
        \sigma^2_{\rho/\rho_0} = f_{\parallel} \sigma^2_{\rho/\rho_0\parallel} + (1 - f_{\parallel})\sigma^2_{\rho/\rho_0\perp}.
    \end{align}
    
    In the high-$\M$ limit we have
    \begin{align}\label{eq:mach_limit}
        \lim_{\M \rightarrow \infty} \sigma^2_{\rho/\rho_0} = f_0 b^2\M_{\rm c}^2 + \frac{1}{2}\left(\sqrt{1 + 8\MaO{2}}-1\right), 
    \end{align}
    which, like the density contrast caused by type~II shocks shown in \S\ref{sec:MHDplanarshock}, is asymptotic to a limit fixed by the magnetic field and turbulent parameters. Ours is the first model to predict such a bound for the total density fluctuations, which is set by the strength of the mean magnetic field, the type of driving, and the volume-filling fraction of subsonic fluctuations travelling along $\vecB{B}_0$. This tells us that even in the high-$\M$ limit the turbulent driving parameter influences the spread of the PDF, frozen into the variance at high-$\M$. 
    Interestingly, the limiting behaviour is independent of $\M$, as the effect of increasingly sharp density contrasts ($\M^2$) of supersonic hydrodynamical shocks is cancelled out by the reduced volume-filling factor of the parallel fluctuations. 
    
    The next limit of interest is the low-$\Mao$ limit, 
    \begin{align}\label{eq:magnetic_limit}
        \lim_{\Mao \rightarrow 0} \sigma^2_{\rho/\rho_0} = f_{\parallel} b^2 \M^2.
    \end{align}
    Thus, the parallel component of the variance is retained in this limit. We can interpret this as hydrodynamical fluctuations are able to survive along the magnetic field but are confined in a small volume, $f_{\parallel}\mathcal{V}$, and hence are reduced by the volume-filling fraction, $f_{\parallel}$. This value will be significantly less than 1 for $\M \gg 10$, as measured in \S\ref{sec:f_par_data}. This is a key distinguishing feature from any of the isotropic magnetised models presented in Equation~\ref{eq:sigma_magnetic} \citep{Molina2012}. Both of these models result in $\sigma^2_{\rho/\rho_0} = 0$ in this limit. 
    
    Finally, in the high-$\Mao$ limit, as the turbulence transitions from magnetised to hydrodynamic,
    \begin{align}\label{eq:hydro_limit}
        \lim_{\Mao \rightarrow \infty} \sigma^2_{\rho/\rho_0} = f_{\parallel} b^2\M^2 + (1 - f_{\parallel}) b^2\M^2 = b^2\M^2,
    \end{align}
    which is the well-known, isotropic, hydrodynamic $\M-\sigma^2_{\rho/\rho_0}$ relation.
    
    \section{Discussion}\label{sec:discussion}
    \subsection{Implications for density fluctuations observed in the ISM}
    We motivated in \S\hbox{\ref{sec:intro}} that there have been a number of ISM observations \citep{Li2011,Li2013,Soler2013,Planck2016a,Planck2016b,Federrath2016,Cox2016,Malinen2016,Tritsis2016,Soler2017,Tritsis2018b,Heyer2020,Pillai2020} and simulations \citep{Soler2017a,Tritsis2018b,Beattie2020,Seifried2020,Kortgen2020,Barreto2021} that show bimodally distributed density structures with respect to the magnetic field.
    
    For example, recent polarimetric observations of nearby MCs reveal that the alignment of filamentary structures in gas column density derived from Herschel submillimetre observations change with the density: high-density structures ($N_{\rm H_2} \propto 10^{23}\,\text{cm}^{-2}$, where $N_{\rm H_2}$ is the number density of the molecular gas column density) tend to be oriented perpendicularly to, and low-density ($N_{\rm H_2} \propto 10^{22}\,\text{cm}^{-2}$) structures parallel to, the mean magnetic field \citep{Planck2016a,Soler2017,Tritsis2018b,Heyer2020,Pillai2020}. Through the analysis of numerical MHD turbulence simulations we have identified some of the physics that causes the bimodal alignment of gas density structures during this density transition. To summarise, the alignment in simulations like ours comes about through an interplay between the divergence (compressibility) and the strain (most likely associated with vortex creation; such as those visualised in Figure~\hbox{\ref{fig:3dPlots}}) in the flow. This has been found in multiple studies of highly-magnetised, compressible plasmas, including those simulations presented in Figure~\hbox{\ref{fig:divDecomp}}, \citep{Soler2017a,Beattie2020,Seifried2020,Kortgen2020}. A different model is developed in \hbox{\citet{Xu2019}}. They use the \hbox{\citet{Goldreich1995}} anisotropy theory to model the extent of low-density filaments in the warm, diffuse and sub-sonic ISM. This model is most likely not applicable for the highly-compressible flows in the supersonic, cool and dense molecular clouds, which need not conform to the incompressible \hbox{\citet{Goldreich1995}} theory. The transition occurs with or without the presence of self-gravity, and hence seems to be a MHD phenomena (however, self-gravitating MHD simulations are shown to enhance the density structures that are perpendicular to $\vecB{B}_0$, e.g. \citealt{Soler2017a,Gomez2018,Barreto2021}). 
    
    Our two-shock model suggests a simple explanation for this density transition described above, irrespective of the physical processes that create the aligned structures, which is beyond the scope of this study. The shock-jump conditions predict that perpendicularly oriented hydrodynamical shocks formed from material flowing along the magnetic field are at least an order of magnitude larger in density contrast compared to parallel oriented fast magnetosonic shocks, formed through shuffling magnetic field lines. We show this qualitatively in Figure~\hbox{\ref{fig:divDecomp}}, with the density contrasts produced by the shocks visualised in the two bottom panels. Hence, we suggest that the observed transition in the density arises from observing these two distinct types of MHD modes. 
    
    A further transition is also now reported at even higher column densities, within the filaments (< 0.1pc) themselves \hbox{\citep{Pillai2020}}. Inside of the filaments gravity-induced accretion gas flows entrain the magnetic field, realigning it with the gas flow. This creates parallel aligned gas channels that accrete onto, for example, hubs between high-density perpendicular oriented filaments. \hbox{\citep{Gomez2018,Pillai2020,Busquet2020}}. 
    
    Pioneering work from \hbox{\citet{Soler2017a}} characterised the bimodality of the first transition in terms of the angle, $\phi$, between $\nabla\rho$ and $\vecB{B}$. They showed that $\cos\phi = \pm 1$ and $\cos\phi = 0$ constitute equilibrium points that an ideal MHD system tends towards. In our work we demonstrated that the physical realisation of this insight corresponds to hydrodynamical shocks that form along $\vecB{B}_0$ ($\cos\phi = 0$) and fast magnetosonic shocks that form perpendicular to $\vecB{B}_0$ ($\cos\phi = \pm 1$). From the perspective of linear MHD wave theory, these are the only two waves that are able to compress the density and form the $\nabla\cdot\mathbf{v} < 0$ structures in the flow. We hypothesise that at least for some sub-Alfv\'enic, supersonic MCs it is these compressible MHD modes that form the over-dense seeds and allow a local region to become Jeans unstable and collapse under gravity.
    
    \subsection{Implications for star formation theory}
    Bottom-up star formation theories treat MCs as the fundamental building blocks that determine galactic star formation rates, and are parameterised by $\sigma_{\rho/\rho_0}^2$, as discussed in \S\ref{sec:intro} \citep{Krumholz2005,Padoan2011,Hennebelle2011,Federrath2012,Federrath2013b,Federrath2015,Mocz2017,Burkhart2018,Hennebelle2019,Lee2019,Lee2020}. Turbulence regulation, in the context of these models, is cast as a battle between density variance, $\sigma_{\rho/\rho_0}^2$, and $(\rho/\rho_0)_{\rm crit}$,  the critical density at which the cloud becomes Jeans unstable and collapses \citep{Federrath2018}. The magnetic field plays a role in these models by reducing the total density variance (as shown in Equation~\ref{eq:sigma_magnetic}), and by introducing additional support through magnetic pressure in the critical density, preventing collapse \citep{Krumholz2005,Federrath2012,Federrath2013b}. However, all of these models treat the magnetic field only as an isotropic contribution via the magnetic pressure. The effects of magnetic tension or the anisotropic effects introduced by a strong guide field are not included in the current theories of star formation. Here we show that the magnetic field, specifically, a sub-Alfv\'enic mean field, which encodes tension into the theory, acts preferentially to suppress small-scale density fluctuations (and hence turbulent fragmentation), preventing $\sigma_{\rho/\rho_0}^2$ from growing beyond a specific limit set by the strength of the field, shown in Equation~\ref{eq:mach_limit} and in the $\ln\rho/\rho_0$-PDFs in Figure~\ref{fig:dens_pdf}. This is an extra form of suppression that $B_0$ has on the star formation rate.
    
    A complete theory for anisotropic star formation would take into account that fluctuations perpendicular to the magnetic field are suppressed significantly, whereas along the field they are not. As such, the theory would predict that star formation may become bimodal with respect to the direction of the large-scale, coherent $\vecB{B}$-field. There are some observational signatures that this may be the case with the star formation rate of some MCs seeming to depend upon the large-scale orientation of the magnetic field \citep{Law2019,Law2020}.
    
    In this study we describe a model for the variance applicable to these types of MCs, but to properly predict the star formation rate in a strongly magnetised environment one also must create an anisotropic model for $(\rho/\rho_0)_{\rm crit}$, which contains both information about the scale in which the turbulence transitions from supersonic to subsonic in rms velocities, i.e., the sonic scale \citep{Federrath2012,Federrath2021}, and the Jeans scale. The morphology of the sonic scale in the presence of a strong magnetic field is unknown, but one can speculate that it most likely will become stretched along the field lines, changing the nature of the critical density and how cloud collapse happens in this regime. What we emphasise here is that there is much work to do in this supersonic, anisotropic, highly-magnetised regime, much of which we intend to pursue in future studies.
    
\section{Summary and key findings}\label{sec:conclusion}
    In this study we explore the density variance, $\sigma_{\rho/\rho_0}^2$, of highly-magnetised, anisotropic, supersonic turbulence across and along the mean magnetic field, $\vecB{B}_0$. In \S\ref{sec:densityPDFs} we discuss the derivation of the $\sigma_{\rho/\rho_0}^2 - \M$ relation and highlight the fundamental connection between shock-jump conditions for the density field and $\sigma_{\rho/\rho_0}^2$. In \S\ref{sec:shockModels}, we describe in detail how we define the geometry of shocks, and finally, we derive our two-shock model for $\sigma_{\rho/\rho_0}^2$. In \S\ref{sec:numerics} we discuss the setup and data processing for the 12~supersonic ($\M > 1$; where $\M$ is the sonic Mach number), trans-/sub-Alfv\'enic ($\Mao \leq 1$; where $\Mao$ is the mean-field Alfv\'enic Mach number) MHD simulations. We show examples of the 2D density slices and the full 3D density fields in Figure~\ref{fig:12_panel} and Figure~\ref{fig:3dPlots}, respectively. In \S\ref{sec:varianceModel} we fit our two-shock density model to the variance data, and in \S\ref{sec:results} we discuss the results from the fit. We summarise the fitting process in the following steps:
    
     \begin{enumerate}
        \item We derive a model for the density variance that takes the general form $\sigma_{\rho/\rho_0}^2 = f_{\parallel}\sigma_{\rho/\rho_0\parallel}^2 + f_{\perp}\sigma_{\rho/\rho_0\perp}^2$, where $\sigma_{\rho/\rho_0\parallel}^2$ comes from type I shocks and $\sigma_{\rho/\rho_0\perp}^2$ comes from type II shocks.
        \item The entire volume of the turbulence must contribute to the total variance, hence we assume that $f_{\parallel} = 1 - f_{\perp}$. This defines a single parameter model $\sigma_{\rho/\rho_0}^2 = f_{\parallel}\sigma_{\rho/\rho_0\parallel}^2 + (1-f_{\parallel})\sigma_{\rho/\rho_0\perp}^2$
        \item We propose a phenomenological model for $f_{\parallel}$, $f_{\parallel} = f_0\left[1 + \left(\frac{\M}{\M_{\rm c}}\right)^4 \right]^{-1/2}$, based on the shock thickness of type I shocks.
        \item We use numerical simulations parameterised by $(\M,\M_{\rm A0})$ to directly measure $\sigma_{\rho/\rho_0}^2$, and calculate $\sigma_{\rho/\rho_0\parallel}^2$ and $\sigma_{\rho/\rho_0\perp}^2$.
        \item Using the numerical data we fit for the two parameters in the $f_{\parallel}$ model, $f_0$, which is associated with the volume-filling fraction of the parallel fluctuations in the subsonic limit, and $\M_{\rm c}$, is the $\M$ when the supersonic flow is significantly saturated with shocks. 
    \end{enumerate}
    
    Finally in \S\ref{sec:discussion} we discuss the implications for interstellar medium structure and magnetised star formation theory. We summarise the key results below:\\
    \begin{itemize}
        \item We derive a model for the density variance, $\sigma^2_{\rho/\rho_0}$, in a sub-Alfv\'enic, supersonic, anisotropic flow regime, where a strong mean magnetic field, $\vecB{B}_0$, creates dynamically different fluctuations parallel and perpendicular to field, characterised by Equation~\ref{eq:2shockModel}. To do this, we generalise the shock-variance relations from \citet{Padoan2011} and \citet{Molina2012}, discussed in \S\ref{sec:shockModels}, partitioning the total volume into two sub-volumes that contain hydrodynamical shocks moving along (parallel to) the magnetic field, and fast magnetosonic shocks moving across (perpendicular to) the field, with details of the orientation and compression mechanism shown in Figure~\ref{fig:shock_schematic}. \\

        \item Our density variance model relies upon the volume-filling fraction, $f_{\parallel}$ -- a measure of how much relative volume the parallel fluctuations occupy along $\vecB{B}_0$. We propose a phenomenological model for $f_{\parallel}$, Equation~\ref{eq:vfrac}, using the shock width described in \citet{Padoan2011}, discussed in detail in \S\ref{sec:volumeFrac}. Using the numerical simulations, we fit our model, with fit parameters listed in Table~\ref{tb:parms}, and illustrated in Figure~\ref{fig:volumeFrac}. By assuming that the parallel and perpendicular fluctuations must occupy the whole volume, we find the parallel fluctuations dominate the volume budget in low-$\M$ flows, whilst the perpendicular fluctuations dominate in high-$\M$ flows, consistent with the compressible structures visualised in Figure~\ref{fig:12_panel} and Figure~\ref{fig:divDecomp}.\\
        
        \item Our new model predicts a finite value of $\sigma^2_{\rho/\rho_0}$ in the high-$\M$ limit, shown in Equation~\ref{eq:mach_limit}, which is set by the strength of $\vecB{B}_0$. This is because a strong $\vecB{B}_0$ field acts preferentially to suppress the small-scale fluctuations introduced in high-$\M$ flows. The new model also predicts a finite value of $\sigma^2_{\rho/\rho_0}$ as $B_0\to\infty$, as shown in Equation~\ref{eq:magnetic_limit}, corresponding to density fluctuations that can persist along $\vecB{B}_0$. In the hydrodynamical limit, as shown in Equation~\ref{eq:hydro_limit}, our model reduces to the well-known relation, $\sigma^2_{\rho/\rho_0} = b^2\M^2$, where $b$ is the turbulent driving parameter. We demonstrate that our variance model provides a good fit to the simulation data in Figure~\ref{fig:varModelFit}. \\
        
        \item In \S\ref{sec:discussion} we discuss how the two different MHD shocks that we use in our model may explain the density transition observed in some nearby MCs. This because the fast magnetosonic shocks, which create density fluctuations parallel to the magnetic field, have density contrasts at least an order of magnitude less than the hydrodynamical shocks, which cause density fluctuations perpendicular to the magnetic field. We also highlight how a strong $\vecB{B}_0$ may additionally suppress star formation by limiting the small-scale density fluctuations, and hence turbulent fragmentation in MCs with $\M \gtrsim 4$.
    \end{itemize}

\section*{Acknowledgements}
    We thank the anonymous reviewer for their detailed review, which increased the clarity and presentation of the study. J.~R.~B.~thanks Shyam Menon for the many productive discussions and acknowledges financial support from the Australian National University, via the Deakin PhD and Dean's Higher Degree Research (theoretical physics) Scholarships, the Research School of Astronomy and Astrophysics, via the Joan Duffield Research Scholarship and the Australian Government via the Australian Government Research Training Program Fee-Offset Scholarship. 
    P.~M.~acknowledges support for this work provided by NASA through Einstein Postdoctoral Fellowship grant number PF7-180164 awarded by the \textit{Chandra} X-ray Centre, which is operated by the Smithsonian Astrophysical Observatory for NASA under contract NAS8-03060.
    C.~F.~acknowledges funding provided by the Australian Research Council (Discovery Project DP170100603 and Future Fellowship FT180100495), and the Australia-Germany Joint Research Cooperation Scheme (UA-DAAD).
    R.~S.~K.~acknowledges financial support from the German Research Foundation (DFG) via the Collaborative Research Center (SFB 881, Project-ID 138713538) 'The Milky Way System' (subprojects A1, B1, B2, and B8). He also thanks for funding from the Heidelberg Cluster of Excellence STRUCTURES in the framework of Germany's Excellence Strategy (grant EXC-2181/1 - 390900948) and for funding from the European Research Council via the ERC Synergy Grant ECOGAL (grant 855130).
    We further acknowledge high-performance computing resources provided by the Australian National Computational Infrastructure (grant~ek9) in the framework of the National Computational Merit Allocation Scheme and the ANU Merit Allocation Scheme, and by the Leibniz Rechenzentrum and the Gauss Centre for Supercomputing (grants~pr32lo, pr48pi, pr74nu).
    The simulation software \textsc{flash} was in part developed by the DOE-supported Flash Centre for Computational Science at the University of Chicago.\\
    
    Data analysis and visualisation software used in this study: \textsc{C++} \citep{Stroustrup2013}, \textsc{numpy} \citep{Oliphant2006,numpy2020}, \textsc{matplotlib} \citep{Hunter2007}, \textsc{cython} \citep{Behnel2011}, \textsc{visit} \citep{Childs2012}, \textsc{scipy} \citep{Virtanen2020},
    \textsc{scikit-image} \citep{vanderWalts2014}. The colour maps used in Figure~\ref{fig:12_panel} and \ref{fig:divDecomp} are sourced from the \textsc{python} package \textsc{cmasher} \citep{Velden2020}.

\section*{Data Availability}
    The data underlying this article will be shared on reasonable request to the corresponding author.


\bibliographystyle{mnras.bst}
\bibliography{March2020.bib} 

\appendix

\section{Flow orientation in the sub-Alfv\'enic, supersonic regime}\label{app:orientation}
\begin{figure}
        \centering
        \includegraphics[width=\linewidth]{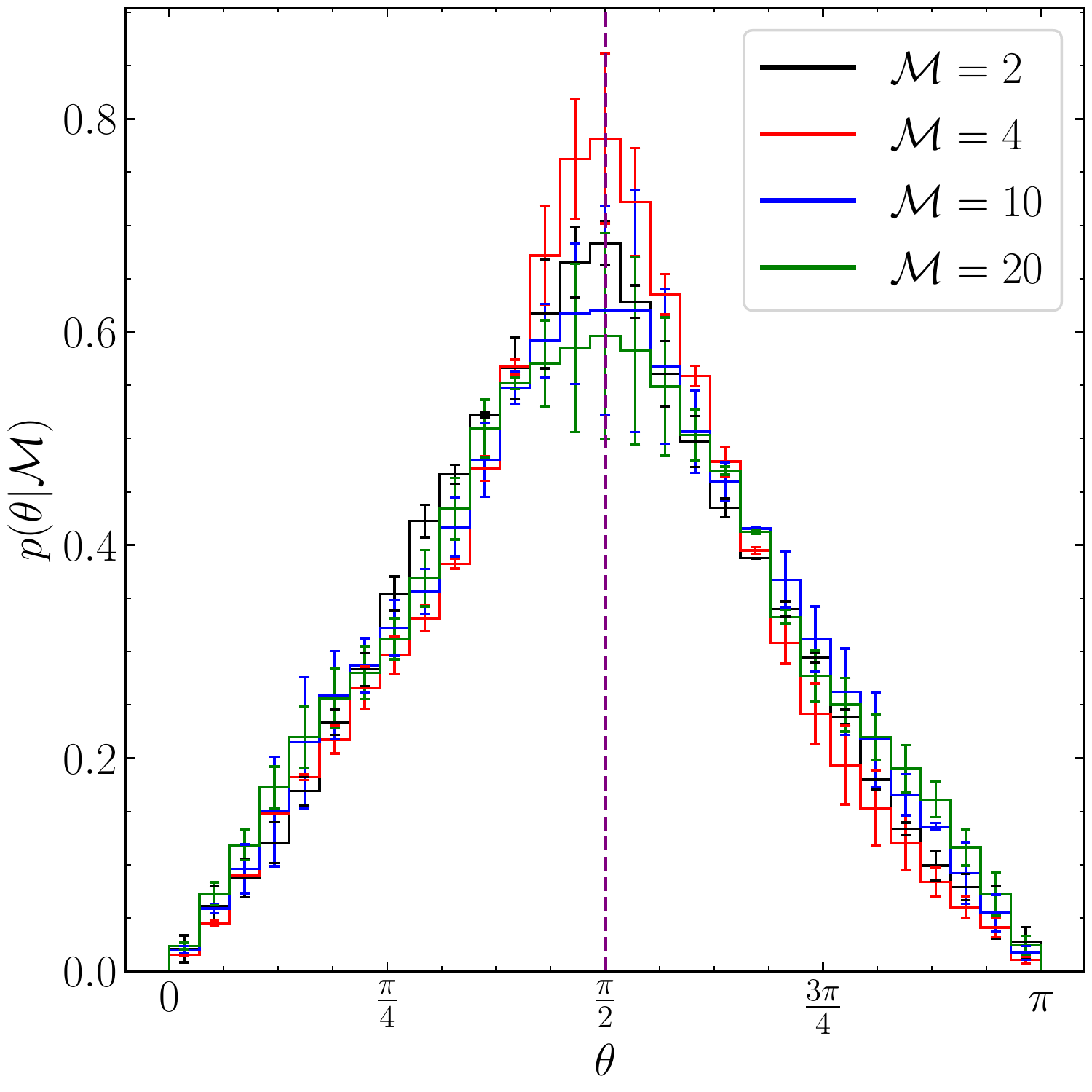}
        \caption{The distribution of the angle, $\theta$, between the local magnetic and velocity field for the ensemble of $\mathcal{M}_{\rm A0}=0.1$ simulations averaged over $5 \leq t/\,T \leq 7$. }
        \label{fig:orientation}
    \end{figure}

    In the sub-Alfv\'enic regime large-scale vortices aligned with $\vecB{B}_0$ form \mbox{\citep{Beattie2020c}}, which arranges the magnetic and velocity fields such that on average $\vecB{v}\perp\vecB{B}$, i.e., $\Exp{\theta} = \Exp{\arccos\left[(\vecB{v}\cdot\vecB{B})/(\|\vecB{v} \| \|\vecB{B}\|)\right]} \approx \pi/2$. We show the full distribution of $\theta$ averaged over $5 \leq t\,T \leq 7$, for our $\Mao = 0.1$ simulations in Figure~\mbox{\ref{fig:orientation}}. We find a highly kurtotic, symmetric distribution centred about $\theta = \pi/2$, as expected for a flow that has vortices rotating about $\vecB{B}_0$ intertwined with intermittent events perturbing $\theta$ away from the mean. The density variance model that we propose in this study applies to this limiting case, where type~I shocks form the intermittent events that perturb $\theta$, and the type~II shocks form by the magnetic field lines being dragged through the supersonic, vortical flow.
    
    Note that in this sub-Alfv\'enic regime there is only a very small fluctuating component compared to the mean-field, $\vecB{B}_0$ (see discussion in \S\mbox{\ref{sec:Bfields}} in the main study). When we compute $\theta$, the angle between $\vecB{B}$ and $\delta\vecB{v}$ in each cell, because $\vecB{B}_0$ dominates the magnetic field, $\vecB{B}\approx\vecB{B}_0$, and hence $\theta \approx \arccos\left[(\vecB{v}\cdot\vecB{B}_0)/(\|\vecB{v} \| \|\vecB{B}_0\|)\right]$. This makes sense for our study since $B_0$ partitions our density domain into parallel and perpendicular components. However, since we are including $\vecB{B}_0$ in our $\theta$, dynamic alignment theory \mbox{\citep{Boldyrev2006,Matthaeus2008}} does not apply, hence we urge caution when comparing Figure~\mbox{\ref{fig:orientation}} with other $\theta$ or $\cos\theta$ distributions between the local magnetic and velocity fields.

\bsp	
\label{lastpage}
\end{document}

%% file: header.tex
\usepackage[T1]{fontenc}
\usepackage{ae,aecompl}

\DeclareRobustCommand{\VAN}[3]{#2}
\let\VANthebibliography\thebibliography
\def\thebibliography{\DeclareRobustCommand{\VAN}[3]{##3}\VANthebibliography}

\usepackage{graphicx}	
\usepackage{amsmath}	
\usepackage{amssymb}	
\usepackage{color}
\usepackage{algorithm}
\usepackage{booktabs}
\usepackage[noend]{algpseudocode}
\usepackage{pifont}
\usepackage{soul}
\usepackage{float}
\usepackage{multibib}
\newcites{Appendix}{Appendix References}
\usepackage{natbib}
\usepackage[flushleft]{threeparttable}
\usepackage{mathrsfs}
\usepackage{widetext}
\usepackage{mathtools}

\graphicspath{{Figures/}}

\newcommand{\appropto}{\mathrel{\vcenter{
  \offinterlineskip\halign{\hfil$##$\cr
    \propto\cr\noalign{\kern2pt}\sim\cr\noalign{\kern-2pt}}}}}

\newcommand{\vecB}[1]{\mathrm{{\bmath{\mathit{#1}}}}}
\newcommand{\Exp}[1]{\mathrm{\left\langle #1 \right\rangle}}

\renewcommand{\d}[1]{\ensuremath{\operatorname{d}\!{#1}}}

\DeclareMathOperator\D{\mathcal{D}}
\DeclareMathOperator\V{\mathcal{V}}
\newcommand{\M}{\mathcal{M}}
\DeclareMathOperator\Ma{\mathcal{M}_{\text{A}}}
\DeclareMathOperator\Mao{\mathcal{M}_{\text{A0}}}
\newcommand{\MaO}[1]{\mathcal{M}^{#1}_{\text{A0}}}

\defcitealias{Brunt2010a}{BFP2010}
\defcitealias{Beattie2020}{BF2020}
\defcitealias{Kolmogorov1941}{K41}
\defcitealias{Menon2020}{MFK2020}
\defcitealias{Hopkins2013}{H2013}

\usepackage{scalerel,tikz}
\usetikzlibrary{svg.path}
\definecolor{orcidlogocol}{HTML}{A6CE39}
\tikzset{orcidlogo/.pic={\fill[orcidlogocol] svg{M256,128c0,70.7-57.3,128-128,128C57.3,256,0,198.7,0,128C0,57.3,57.3,0,128,0C198.7,0,256,57.3,256,128z}; \fill[white] svg{M86.3,186.2H70.9V79.1h15.4v48.4V186.2z} svg{M108.9,79.1h41.6c39.6,0,57,28.3,57,53.6c0,27.5-21.5,53.6-56.8,53.6h-41.8V79.1z M124.3,172.4h24.5c34.9,0,42.9-26.5,42.9-39.7c0-21.5-13.7-39.7-43.7-39.7h-23.7V172.4z} svg{M88.7,56.8c0,5.5-4.5,10.1-10.1,10.1c-5.6,0-10.1-4.6-10.1-10.1c0-5.6,4.5-10.1,10.1-10.1C84.2,46.7,88.7,51.3,88.7,56.8z};}}
\newcommand\orcidicon[1]{\href{https://orcid.org/#1}{\mbox{\scalerel*{
\begin{tikzpicture}[yscale=-1,transform shape]\pic{orcidlogo};
\end{tikzpicture}}{|}}}}